\providecommand{\columnwiselinenumbersfalse}{}%
\providecommand{\runningpagewiselinenumbers}{}%
\PassOptionsToPackage{desactivate}{linenoaa}
\documentclass{aa}  

\usepackage{graphicx}
\usepackage{txfonts,textcomp}
\usepackage{mathrsfs}
\usepackage{booktabs}
\usepackage[svgnames]{xcolor}
\usepackage{hyperref}

\usepackage{orcidlink}
\definecolor{C0blue}{RGB}{31,119,180}
\definecolor{C1orange}{RGB}{255,127,14}
\definecolor{C2green}{RGB}{44,160,44}
\definecolor{C3red}{RGB}{214,39,40}
\definecolor{C4purple}{RGB}{148,103,189}
\definecolor{C5brown}{RGB}{140,86,75}
\definecolor{C6pink}{RGB}{227,119,194}
\definecolor{C7gray}{RGB}{127,127,127}
\definecolor{C8olive}{RGB}{188,189,34}
\definecolor{C9cyan}{RGB}{23,190,207}

\newcommand{\reddtext}[1]{\textcolor{DarkRed}{#1}}

\hypersetup{
    breaklinks=true,
    colorlinks=true,
    linkcolor=C2green,
    citecolor=C0blue,
    filecolor=magenta,      
    urlcolor=C9cyan,
    pdftitle={EMPEROR - Pe{\~n}a, P. and Jenkins, J},
    pdfauthor={Pe{\~n}a, Pablo},
    }

\newcommand{\refsec}[1]{\hyperref[#1]{\S\ref*{#1}}}
\newcommand{\refeq}[1]{\hyperref[#1]{Eq. \ref*{#1}}}
\newcommand{\reffig}[1]{\hyperref[#1]{Fig.\ref*{#1}}}
\newcommand{\reftab}[1]{\hyperref[#1]{Table \ref*{#1}}}
\newcommand{\refapp}[1]{\hyperref[#1]{Appendix \ref*{#1}}}

\newcommand{\ms}{ms$^{-1}$}

\newcommand{\me}{M$_{\rm{\oplus}}$}

\newcommand{\emp}{\texttt{EMPEROR}}
\newcommand{\emc}{\texttt{EMCEE}}
\newcommand{\reddemc}{\texttt{reddemcee}}
\newcommand{\dynesty}{\texttt{dynesty}}

\newcommand{\ins}{\text{INS}}

\newcommand{\Uniform}[2]{$\sim\mathcal{U}(#1, #2)$}
\newcommand{\Normal}[2]{$\sim\mathcal{N}(#1, #2)$}
\newcommand{\LNormal}[2]{$\sim\mathcal{LN}(#1, #2)$}

\newcommand{\gprot}{$\mathcal{GP}_{\mathrm{rot}}$}
\newcommand{\gpsho}{$\mathcal{GP}_{\mathrm{SHO}}$}
\newcommand{\gpgrot}{$\mathcal{GP}_{\mathrm{grot}}$}
\newcommand{\gpthreesho}{$\mathcal{GP}_{\mathrm{3SHO}}$}
\newcommand{\gprotplussho}{$\mathcal{GP}_{\mathrm{rot+SHO}}$}
\newcommand{\gptworot}{$\mathcal{GP}_{\mathrm{2rot}}$}

\newcommand{\lnz}{$\ln{\mathcal{Z}}$}

\newcommand{\lnzerr}{$\hat{\sigma}_{\mathcal{Z}}$}

\newcommand{\prot}{$P_{\text{rot}}$}
\newcommand{\pmag}{$P_{\text{mag}}$}

\newcommand{\juliet}{\texttt{Juliet}}

\begin{document}

   \title{EMPEROR}

   \subtitle{I. Exoplanet MCMC parallel tempering for RV orbit retrieval}

   \author{Pablo A. Pe{\~n}a R.\orcidlink{0000-0002-8770-4398}\inst{1,2} \and James S. Jenkins\orcidlink{0000-0003-2733-8725}\inst{1,2}}

   \institute{Instituto de Estudios Astrof\'isicos, Facultad de Ingenier\'ia y Ciencias, Universidad Diego Portales, Av. Ej\'ercito 441, Santiago, Chile \\
   \and Centro de Astrof\'isica y Tecnolog\'ias Afines (CATA), Casilla 36-D, Santiago, Chile\\}

   \date{}

\abstract 
{}
{This paper presents \emp, an open-source Python-based framework designed for the efficient detection and characterisation of exoplanets by using radial velocity (RV) methods. Its combination of performance, flexibility, and ease of use makes it a robust tool for any exoplanet detection endeavour. \emp~integrates dynamic nested sampling (DNS) and adaptive parallel tempering (APT) Markov chain Monte Carlo (MCMC) techniques, supporting multiple noise models such as Gaussian processes (GPs) and moving averages (MA). The framework facilitates systematic model comparison using statistical metrics, including Bayesian evidence (\lnz) and Bayesian information criterion (BIC), while providing automated, publish-ready visualisations.}
{\emp~was evaluated  across three distinct systems to assess its capabilities in different detection scenarios. The sampling performance, model selection, and  search for Earth-mass planets were investigated in data for 51 Pegasi, HD 55693, and Barnard's star (GJ 699).}
{For 51 Pegasi, we find APT achieves an effective sampling increase by a factor of 3.76 over DNS, while retrieving tighter parameter estimates. For HD 55693, the stellar rotation, \prot=$29.72^{+0.01}_{-0.02}$, and magnetic cycle, \pmag=$2557.0^{+70.1}_{-36.7}$, were recovered, while demonstrating the sensitivity of \lnz~to prior selection. For Barnard's star, several noise models were compared and the confirmed planet parameters were successfully retrieved with all of them. The best model shows a period of 3.1536$\pm$0.0003~d, minimum mass of 0.38$\pm$0.03~\me, and semi-major axis of 0.02315$\pm$0.00039~AU. }
{Purely statistical inference might be insufficient on its own for robust exoplanet detection. Effective methodologies must integrate domain knowledge, heuristic criteria, and multi-faceted model comparisons. The versatility of \emp~in handling diverse noise structures, its systematic model selection, and its improved performance make it a valuable tool for RV exoplanetary studies.}


\maketitle

\section{Introduction}  \label{sec:intro}

Within the realm of  planetary systems orbiting stars other than the Sun, there appears to be an almost unlimited number of plausible configurations. The field of exoplanet research is still relatively young; however, the diversity of planetary systems is notably high (e.g. \citealp{1995Natur.378..355M,2006MNRAS.369..249J,2012ApJ...752....1R,2013A&A...556A.126A, 2017Natur.542..456G}, and many others).

Detection is the first step in fully comprehending the nature of planetary systems observationally. Novel methods are continually being developed to uncover systems that are difficult to detect, such as very low-mass planets, multi-planet systems, wide-orbiting companions, or systems with unusual architectures. Numerous new techniques have been proposed that extend beyond the traditional approach of utilising Lomb-Scargle periodograms \citep[LSP, ][]{1976Ap&SS..39..447L, 1982ApJ...263..835S}. As such, newer periodogram schemes, as well as new algorithms, have been developed, allowing for more information to be extracted from the data, such as  generalised LSP \citep{2009A&A...496..577Z}, minimum mean squared error \citep{2014MNRAS.441.2253J}, maximum likelihood periodograms \citep{2013MNRAS.429.2052B}, Gaussian processes \citep{2015MNRAS.452.2269R}, deep learning \citep{2018AJ....155...94S}, and many others (see \citealp{2016arXiv161100766S} for a short overview).

The application of Bayesian modelling under a parallel tempering Markov chain Monte Carlo (MCMC) framework \citep{2007MNRAS.381.1607G}, particularly when combined with correlated-noise modelling, has recently made strides in the signal detection arena, including the hunt for exoplanets. These methods have led to the discovery of small planet candidates orbiting Sun-like stars \citep{2013A&A...556A.111T, 2013ApJ...766...67J} and multi-planet systems orbiting small M dwarf stars (\citealp{2016Natur.536..437A, 2017Natur.542..456G}); in known systems, they have improved constraints on planet candidates or revealed new ones \citep{2017MNRAS.470.4794F}. In addition, these methods have helped to disprove or revise previous claims, such as those for $\alpha$ Centauri B b \citep{2012Natur.491..207D, 2013ApJ...770..133H, 2015MNRAS.452.2269R}, GJ 581 d and g \citep{2010ApJ...723..954V, 2015ApJ...805L..22R}, or Kapteyn b \citep{2014MNRAS.443L..89A, 2015ApJ...805L..22R}.

Despite these advances, the application of MCMC methods in RV inference presents key challenges, such as the prolonged computing time required to reach an optimal solution; the multi-modal nature of Keplerian signals, which makes it difficult to ensure that the global posterior maximum has been identified; and the difficulty in achieving full automation in the analysis.

Several radial velocity (RV) fitting frameworks address different parts of this landscape; for example, \texttt{RadVel} \citep{2018PASP..130d4504F}, \juliet~\citep{2019MNRAS.490.2262E}, \texttt{kima} \citep{2018JOSS....3..487F}, and \texttt{exostriker} \citep{2019ascl.soft06004T}. Whilst these provide flexibility in planet detection and parameter estimation, in this work, we introduce \emp\,(Exoplanet Mcmc Parallel tEmpering for Rv Orbit Retrieval). This is a Python-based, modular code that offers a single automated workflow that i) natively employs an affine-invariant APT MCMC engine with built-in evidence estimation; ii) supports modular per-instrument noise and stellar activity modelling, and optional dynamical-stability priors for multi-planet solutions; and iii) pairs an auto-compiled, parallelised likelihood back-end with publish-ready diagnostics, enabling reliable exploration of broad parameter spaces with minimal manual intervention.

Internally, the posterior is sampled using an APT implementation built on the affine-invariant ensemble sampler, \reddemc\,\citep{reddemcee-paper}. Alternative sampling methods, \emc\,\citep{2013PASP..125..306F}, PyMC3 NUTS sampler \citep{2015arXiv150708050S}, and \dynesty\,\citep{2020MNRAS.493.3132S}, are also available; however, the empirical tests conducted in this study indicate that the APT approach can provide greater confidence in identifying high-probability modes of multi-modal posteriors, subject to the usual caveat that chains substantially exceed the integrated autocorrelation time.

The code supports multi-instrument and multi-planet analysis, as well as various noise models. It conducts automated model comparison to identify and return the optimal model that best describes the data. \emp~has already been used in published works to detect new planet candidates and confirm existing ones \citep[e.g.][]{2017AJ....153...51W, 2019MNRAS.487..268J, 2020NatAs...4..419B, 2022A&A...664A..94P, 2023MNRAS.518.2627V}.

This work describes \emp's functionality and the remainder of this manuscript is structured as follows. In \hyperref[sec:bayesian_framework]{\S\ref{sec:bayesian_framework}, we} outline the Bayesian framework. In \hyperref[sec:code]{\S\ref{sec:code}, we} explain the structure and configuration of \emp, including automated pre-processing and post-processing procedures. In \hyperref[sec:benchmarks]{\S\ref{sec:benchmarks}, we} present validation tests and benchmarks using three representative cases:\ 1) 51 Pegasi, enabling a direct comparison with published results to evaluate both performance and efficiency of \emp; 2) HD 55693, a system where stellar activity induces RV signals, used to test model comparison under correlated noise; and 3) Barnard's star, which explores modelling nuances in current ultra-precise data. In \hyperref[sec:discussion]{\S\ref{sec:discussion}, we} discuss the complexities of modelling and the advantages of using \emp~in any RV signal detection endeavour. Finally, \hyperref[sec:conclusions]{\S\ref{sec:conclusions}} concludes with an overview of the analysis performed in this paper.

\section{Bayesian framework}  \label{sec:bayesian_framework}

Early RV searches relied on frequentist fitting techniques such as chi-squared minimisation and LSP \citep{1976Ap&SS..39..447L, 1982ApJ...263..835S}. Coupled with local optimisers such as the Levenberg–Marquardt algorithm \citep{levenberg44, marquardt63}, these approaches work well for strong isolated signals, but falter when likelihood surfaces became multi-modal, which is a characteristic of multi-planet systems or low signal-to-noise ratio (S/N) data. Exploring complex Keplerian architectures or alternative noise models quickly became computationally prohibitive, making Earth-mass planets difficult to detect.

Bayesian inference overcomes these hurdles by mapping the full posterior probability of the parameters rather than converging on a single best-fit. MCMC samplers draw from that posterior, naturally quantifying parameter uncertainties \citep{2007MNRAS.381.1607G, 2013A&A...551A..79T}. Affine invariant samplers \citep{2010CAMCS...5...65G, 2012ApJ...745..198H} are particularly effective in the high-dimensional, highly correlated parameter spaces typical of RV studies.

\subsection{The Bayesian approach} \label{sec:bayesian_framework_bayesian_approach}

Bayes' theorem expresses the posterior, $p(\pmb{\theta} | D, M),$ of model parameters, $\pmb{\theta}$, given a set of data, $D$, and model, $M$, as:

\begin{equation} \label{eq:bayes_theorem}
    p(\pmb{\theta} | D, M) = \frac{p(\pmb{\theta} | M) \cdot p(D | \pmb{\theta}, M)}{p(D | M)}
,\end{equation}

where the numerator combines the prior $p(\pmb{\theta} | M)$ with the likelihood $p(D | \pmb{\theta}, M)$ and the denominator $p(D | M)$ is the Bayesian evidence that normalises the posterior \citep{robert_2007, 2019arXiv190508737F, 2022arXiv220211678L}. Computing this evidence later enables objective model comparison via Bayes factors.

In RV analysis the deterministic component of the signal is a sum of Keplerian curves,

\begin{equation} \label{eq:keplerian}
    \mathcal{K}(t) = \sum_{j=1}^{N_{\mathrm{pl}}} K \cdot [\cos(\nu_j(t) + \omega_j) + e_j\cos(\omega_j)]
,\end{equation}

where each planet $j$ is described by the tuple ($P_j$, $K_j$, $e_j$, $\omega_j$, $T_{0,j}$) and $\nu_j$ is the true anomaly. Additional terms such as instrumental offsets, long-term trends, or linear correlations with stellar activity indices, can be used to expand \refeq{eq:keplerian} without altering its basic form.

The data consist of $N$ RV measurements ({$\rm{RV}_i$,$\sigma_i$,$t_i$), optionally accompanied by activity indices. Assuming independent
Gaussian errors, augmented by an instrument-specific `jitter',
$\sigma_{\mathrm{INS}}$, the likelihood is

\begin{equation}\label{eq:likelihood}
    p(D | \pmb{\theta}) = \prod_{\ins}\prod_i^N \sqrt{\frac{1}{2 \pi (\sigma_i^2 + \sigma_{\ins}^2)}} \exp{(- \frac{\xi_i^2}{2 (\sigma_i^2 + \sigma_{\ins}^2)})}
,\end{equation}

with residuals of $\xi_{i,\mathrm{INS}}\!=\!{\rm RV}_{i,\mathrm{INS}} -
K(t_i) - \Gamma_{\mathrm{INS}}(t_i) - A_{\mathrm{INS}}(t_i)$, where
$\Gamma$ captures offsets and trends and $A$ any linear
activity terms.  More sophisticated noise models (moving average, red noise, Gaussian processes, etc.) replace or expand the simple white-noise jitter, but fit seamlessly into the same Bayesian machinery.

\subsection{Model proposal} \label{sec:bayesian_framework_model_proposal}

A complete RV model combines one or more Keplerian signals with terms that capture instrumental systematics and stellar variability.
Each planet contributes the familiar five parameters
$(P,\,K,\,e,\,\omega,\,\phi)$, namely, the period, semi-amplitude, eccentricity, longitude of periastron, and phase of periastron passage, so that the baseline dimensionality is $5N_{\mathrm{pl}}$. Nonetheless, \hyperref[eq:keplerian]{Eq. \ref{eq:keplerian}} is agnostic to choice of angular variables, and different parameterisations are supported in \emp, for example, the textbook $(P,\,K,\,\phi,\,e,\,\omega)$ \citep{2018exha.book.....P}. The equivalent set (P, $K_c$, $K_s$, $e_c$, $e_s$) was proposed by \citet{2012ApJ...745..198H}, where 

\begin{equation} \label{eq:hou_cv_full}
\begin{aligned}
    K_c &= \sqrt{K}\cos(\phi),          &   K_s &= \sqrt{K}\sin(\phi), \\
    e_c &= \sqrt{e}\cos(\omega),  &   e_s &= \sqrt{e}\sin(\omega).
\end{aligned}
\end{equation}

This re-parameterisation is used to linearise the circular parameters $\omega$ and $\phi$, improving sampler performance. In photometric studies, the time of inferior conjunction, $T_0$, is often preferred over the phase, $\phi$. The phase-space dependence on $T_0$ with $P$ makes the circularisation less efficient given the fact that the Jacobian for this change of variable will be different from unity. Consequently, the set ($P$, $K$, $T_0$, $e_c$, $e_s$) is recommended for such applications.

Using the mean anomaly as the angular coordinate is discouraged, as it is tied to an arbitrary epoch, $t_0$. This complicates any interpretation and, hence, the uninformative priors as well.

Offsets and long-term accelerations can be modelled with the polynomial, $\Gamma$,

\begin{equation} \label{eq:acceleration}
    \Gamma_{i, \ins}(t_i, a) = \sum{\frac{\partial^a \gamma}{\partial^a dt} \cdot t^a}
,\end{equation}

 where $a$=$0$ returns a constant offset, $a$=$1$, yields a linear term, $a$=$2$ a quadratic term, and so on.

First-order linear correlations with activity indices $A_{i,\ins}$, which could be chromospheric stellar activity proxies (e.g. the $S$-index or log$R'_{\rm{HK}}$, see \citealp{2008A&A...485..571J,2011A&A...531A...8J}) or line asymmetry measurements (e.g. full width at half maximum, bisector index slope; see \citealp{2001A&A...379..279Q}) can be modelled via
 
\begin{equation}  \label{eq:stellar-activity-corr}
    A_{i,\ins} = \sum_{\mathcal{A}} \mathcal{C}_{\mathcal{A},\ins} \cdot \mathcal{A}_{i,\ins}
,\end{equation}

where $\mathcal{A}_{i,\ins}$ denotes each measured activity index and $\mathcal{C}_{\mathcal{A},\ins}$ is the corresponding coefficient.

Short-timescale, colour-correlated noise is captured by a
$q$-order moving average,

\begin{equation}  \label{eq:moving-average}
    R_{i,\ins} = \sum_{q}\Phi_{\ins,q} \exp(\frac{-|t_{i-q}-t_i|}{\tau_{\ins,q}}) \xi_{i-q,\ins}
,\end{equation}
where $\Phi$ and $\tau$ are the correlation strength and decay timescale, respectively.

With $\ins$ counting independent data sets, the number of dimensions of the model is given by $5N_{\mathrm{pl}} + 2 (\ins + q) + a + \mathcal{A}$.

\subsection{Extended stochastic modelling}

\emp~supports custom modelling options designed to capture low-frequency modulations in stellar signals. Two examples are the single sinusoid (\hyperref[eq:sinusoid-model]{Eq. \ref{eq:sinusoid-model}}) and the 'half-period double sinusoid' (\hyperref[eq:magnetic-cycle]{Eq. \ref{eq:magnetic-cycle}}), both of which are used to approximate magnetic cycles or other quasi-periodic behaviours,

\begin{align}
  S_{i}(t) &= K_i \cos{(\omega_i t + \phi_i),} \label{eq:sinusoid-model}\\
  M(t) &= K_1 \cos{(\omega_1 t + \phi_1)} + K_2 \cos{(2\omega_1 t + \phi_2)} \label{eq:magnetic-cycle}.
\end{align}

For complex or unknown systematics, \emp~supports Gaussian processes \citep{2006gpml.book.....R} via the \texttt{celerite} library \citep{2017AJ....154..220F}. GPs are frequently used to model stellar rotation signals, given their quasi-periodic nature. \emp~provides two distinct pre-configured GP kernels to do so. Both are based on mixtures of simple harmonic oscillator (SHO) kernels. The power spectral density of each SHO component is defined by

\begin{equation}  \label{eq:gp-sho-psd}
    S(\omega) = \sqrt{\frac{2}{\pi}} \frac{S_0 \omega_0^4}{(\omega^2-\omega_0^2)^2+\omega^2\omega_0^2/Q^2}
,\end{equation}

where $\rho=2\pi/\omega_0$ is the undamped period, $Q$ is the quality factor (related to the damping timescale $\tau=2Q/\omega_0$), $S_0$ is the amplitude of the process, and $\omega_0$ is the undamped angular frequency.

$\mathcal{GP}_{\mathrm{rot}}$($\sigma$,$\rho$,$Q_0$,$dQ$,$f$) follows \citet{2018RNAAS...2...31F} and comprises two SHO components at periods, $\rho$ and $\rho/2$, effectively capturing the fundamental rotation frequency and its first harmonic. Five parameters control their amplitudes and damping,

\begin{equation} \label{eq:gprot}
\begin{aligned}
    \rho_1 &= \rho,           & Q_1 &= \tfrac{1}{2} + Q_0 + \delta Q,    & S_1 &= \frac{\sigma^2}{(1+f)\omega_1 Q_1}, \\
    \rho_2 &= \frac{\rho}{2}, & Q_2 &= \tfrac{1}{2} + Q_0,               & S_2 &= \frac{f\sigma^2}{(1+f)\omega_2 Q_2}.
\end{aligned}
\end{equation}

The second kernel, $\mathcal{GP}_{\mathrm{grot}}$($\rho$,$\tau$,$A_1$,$A_2$), uses support variables $A_1$ and $A_2$ instead of $S_1$ and $S_2$ directly, and shares the same overall structure as the first kernel; however, it enforces a common damping timescale, $\tau$, for both SHO components,

\begin{equation} \label{eq:gp-grot}
\begin{aligned}
    \rho_1 &= \rho,              & Q_1 &= \frac{\pi\tau}{\rho_1},              & S_1 &= \frac{A_1}{2\tau} \left(\frac{\rho_1}{\pi}\right)^2, \\
    \rho_2 &= \frac{\rho}{2},    & Q_2 &= \frac{\pi\tau}{\rho_2},              & S_2 &= \frac{A_2}{2\tau} \left(\frac{\rho_2}{\pi}\right)^2.
\end{aligned}
\end{equation}

Both kernels offer a flexible way to model spot-modulated stellar rotation and the choice between them depends on the specific system’s characteristics or on prior constraints for the damping timescale, the rotation period, or the amplitude.

\section{\emp ~code}  \label{sec:code}

\emp~is a modular framework that automates RV planet searches with minimal hand-tuning. Three high-level stages, pre-processing, run compilation, and post-processing are executed in a loop that can optionally add model `blocks' (i.e. complexity) and re-launch itself (\hyperref[fig:code-flowchart]{Fig. \ref{fig:code-flowchart}}).
At its core lies \reddemc, the APT sampler, backed by alternative engines (\emc, \dynesty, \texttt{PyMC3}) that can be swapped in through a single keyword.

Modelling is done via `blocks', which are Python objects that bundle 1) a computational model, 2) parameter-holding objects `specs', and 3) metadata such as dimensionality or \LaTeX~representation. A typical single-planet fit (1K) would include a Keplerian block plus offset, and jitter blocks. Because every block carries its own solver, priors, and script-writing routine, new physics can be added (or removed) via one-line flags.

\begin{figure}
    \includegraphics[width=\columnwidth]{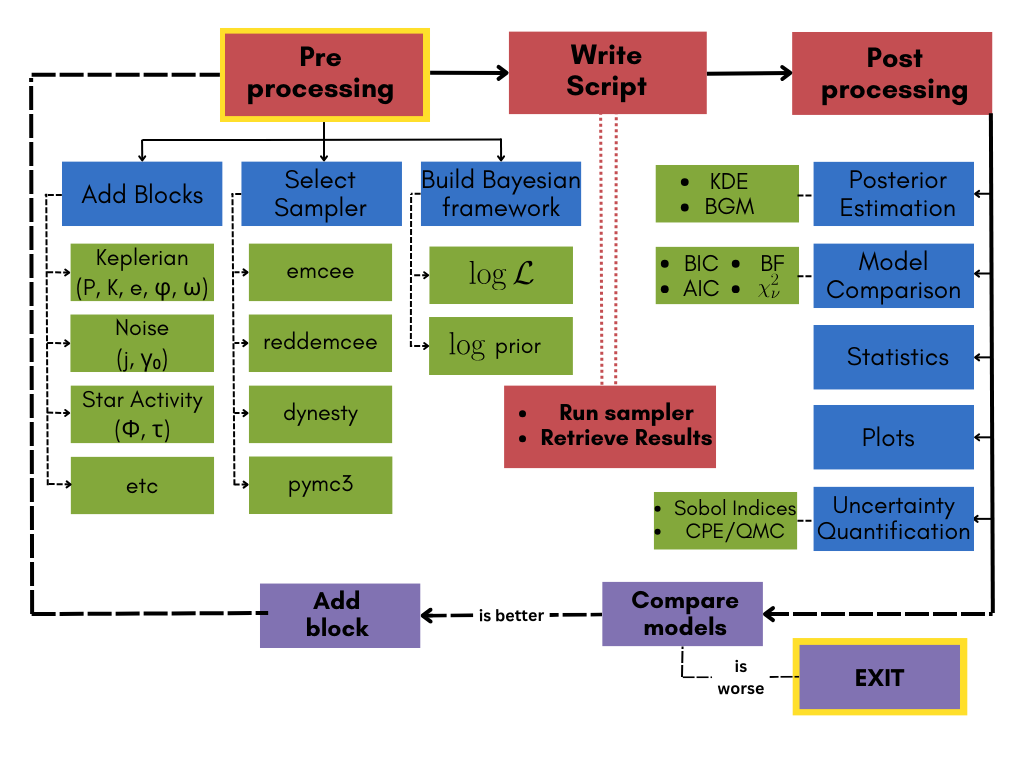}
    \caption{Three distinct steps of the code, shown in red: pre-processing, where user inputs and data are read; run compilation, where the code is executed; and post-processing, where statistical measures are calculated. The purple blocks with dashed arrows represent the optional step of repeating the loop. Blue highlights the main stages of each step, and available alternatives are displayed in green.}
    \label{fig:code-flowchart}
\end{figure}

This section describes the inner workings of the code and provides guidance on its usage. The APT sampler, \reddemc\footnote{\url{https://github.com/ReddTea/reddemcee}}, the default custom Kepler equation solver written in \texttt{cython}\footnote{\url{https://pypi.org/project/fast-kepler}}, which natively admits different parameterisations, and \emp\footnote{\url{https://github.com/ReddTea/astroemperor}} are publicly available, open source, and hosted on GitHub;  they can be easily installed via \texttt{pip}.

\subsection{Stage 1:\ Preprocessing} \label{sec:code_preprocessing}

\paragraph{Sampler selection:} \reddemc~is an APT MCMC sampling algorithm, designed to efficiently handle highly multi-modal posteriors, and serves as the core engine of \emp. Any native options for the chosen sampler are supplied in a dictionary that \emp~passes to the secondary script. The available samplers (\reddemc, \texttt{emcee}, \dynesty, and \texttt{PyMC3}) are briefly explained in \hyperref[sec:appendix_samplers]{Appendix \ref{sec:appendix_samplers}}.

\paragraph{Data ingestion:} Each input file is treated as a separate instrument; RVs are mean-subtracted to centre offsets. Columns beyond (time, RV, error) are parsed as activity indices and automatically mapped to the stellar activity  block.

\vspace{-\baselineskip}
\paragraph{Default Bayesian priors:}
Most parameters employ flat (uniform) priors, whereas targeted normal priors are adopted where experience shows it improves convergence (i.e. jitter, eccentricity). 
In \emp, each parameter's prior is attached to its corresponding spec, storing the functional form of the prior and any relevant hyper-parameters. This metadata is then hard-coded into a temporary script generated during runtime and each of these functions and variables can be adjusted in the code's inputs.

\vspace{-\baselineskip}
\paragraph{Modelling:}
Single-line switches activate moving-average noise, Gaussian process kernels, alternative Keplerian parameterisations (as described in \hyperref[sec:bayesian_framework_model_proposal]{\S\ref{sec:bayesian_framework_model_proposal}}), and more.

\subsubsection{Default Bayesian priors} \label{sec:code_default_bayesian_priors}

\paragraph{Period:} a uniform prior, where $P_{\mathrm{min}}$ is 0.1 days and $P_{\mathrm{max}}$ is three times the baseline of the RV data. A common heuristic requires at least one full cycle of data coverage; however, extending the upper bound to three times the observation baseline can reveal long-period signals that are detectable with LSPs or Bayesian approaches, although weakly constrained (which could be crucial for disentangling multiple signals).

\vspace{-\baselineskip}
\paragraph{Semi-amplitude:} a uniform prior, with $K_{\min}$ = 0.1 $(ms^{-1})$ and a conservative $K_{\max}$ = 3 $\cdot$ $F$, where $F$ corresponds to the RV value that is furthest from the RV mean, F = $\max(|\mathrm{RV}|)$. Although some practitioners choose to use three times the RMS of the RVs as $K_{\max}$, experience shows that in certain cases (e.g. high amplitude signals with long observational gaps from multiple instruments) large offsets can mimic high amplitudes.

\vspace{-\baselineskip}
\paragraph{Eccentricity:} a truncated normal prior, $\mathcal{N}(0, \sigma_{e}^2)$ with $\sigma_e$=0.3 as the code's default, where $\Pi(e<0)$=0. This choice reflects the overall observed distribution of planetary eccentricities, penalising higher eccentricities but allowing them if supported by the data \citep[see][]{2013A&A...551A..79T}. Another common choice for the eccentricity prior is the Beta distribution $\sim \beta(0.867, 3.03)$ \citep[see][]{2013MNRAS.434L..51K}.

\vspace{-\baselineskip}
\paragraph{Offsets and acceleration:} uniform priors in the intervals $[-F, F]$ and $[-1, 1]$ for $\gamma$ and $\dot{\gamma}$, respectively, where $F$ denotes the full RV coverage.

\vspace{-\baselineskip}
\paragraph{Jitter:} truncated normal prior, where the default choice is $\mu_{\sigma}$=5ms$^{-1}$, and $\sigma_{\sigma}$=5ms$^{-1}$. To avoid the jitter from either blowing up or collapsing to zero, while reflecting typical expectations for the star, it is sensible to use a broad Gaussian prior. These values should be adapted according to the instrument precision or stellar type. For example, when searching for low-mass planets in quiescent stars, a lower mean value is favoured \citep{2005PASP..117..657W,2006MNRAS.372..163J,2011ApJ...734...70A}.

\vspace{-\baselineskip}
\paragraph{Moving average:} uniform priors, for the MA Coefficient $\Phi\in[-1, 1]$, to ensure a stationary process that does not arbitrarily diverge over time \citep{2013A&A...549A..48T}, and for the MA Timescale $\tau \in [0.1, 10]$ was chosen per \citet{2016Natur.536..437A}. For $\Phi$, another common choice is $\pm$0.99, avoiding edge behaviour near $\pm1$, circumventing long auto-correlation tail effects.

\begin{table}
    \caption{\label{tab:prior_choices} Default \emp~parameter priors.}
    \centering
        \begin{tabular}{llll}
        \toprule
        Parameter   & Prior & Parameter & Prior \\
        \midrule
        \midrule
        Keplerian    & & Others &   \\
        \midrule
        P (days)        & $\mathcal{U}(0.1, 3 \cdot \max(t))$  &$\gamma$ (\ms)    & $\mathcal{U}(-F, F)$      \\
        K (\ms)   & $\mathcal{U}(0.1, 3 \cdot F)$ & $\theta_{\ins}$ (\ms)  & $\mathcal{N}(5, 5^2)$ \\
        $\phi$ (rads)   & $\mathcal{U}(0, 2\pi)$  & MA $\Phi$                 & $\mathcal{U}(-1, 1)$      \\
        $\omega$ (rads) & $\mathcal{U}(0, 2\pi)$  & MA $\tau$ (days)& $\mathcal{U}(0, 10)$      \\
        e               & $\mathcal{N}(0, 0.3^2)$ & $\dot{\gamma}$ (\ms/yr)   & $\mathcal{U}(-1, 1)$      \\

        \bottomrule
        \end{tabular}
    \tablefoot{$\mathcal{N}$ and $\mathcal{U}$ stand for normal and uniform distribution, respectively.}
    
\end{table}

\subsubsection{Additional priors: Dynamical stability} 
 \label{sec:code_additional_priors_stability}

Beyond the standard prior choices (summarised in \hyperref[tab:prior_choices]{Table \ref{tab:prior_choices}}), there are additional priors that can be used. A dynamical stability prior can be applied for systems with two or more Keplerian signals, to allow only Hill stable configurations to be sampled. Following \cite{1993Icar..106..247G} , we can compute a first-order approximation to the \cite{1982CeMec..26..311M} Hill Criterion,

\begin{equation} \label{eq:hill_criterion}
\begin{aligned}
    \alpha^{-3}(\mu_1 + \frac{\mu_2}{\delta^2})(\mu_1\gamma_1 - \mu_2\gamma_2\delta)^2 > 1 + \mu_1\mu_2(\frac{3}{\alpha})^\frac{4}{3};\\
    \alpha = \mu_1 + \mu_2;\qquad \gamma_j = \sqrt{1 - e_j^2};\qquad \delta = \frac{a_2}{a_1}.
\end{aligned}
\end{equation}

Here, $\mu_j = \frac{m_j}{M}$, with $M$ the stellar mass, $m_j$ the minimum-mass, $a_j$ the semi-major axis, $e_j$ the eccentricity, and $j$=1,2 denotes each planet, with subscript 1 standing for the inner orbit.

Alternatively, since the Hill criterion is less precise for resonant systems, a dynamical stability prior based on the AMD framework \citep{2018A&A...617A..93P, 2019ESS.....420006L} is available. With $\gamma \equiv \frac{m_1}{m_2}$ and $\alpha \equiv \frac{a_1}{a_2}$ the semi-major axis ratio, for a small enough ratio of planetary masses with the host star, $\varepsilon$, given that the relative AMD $\mathscr{C}$ verifies the inequality,

\begin{equation}
        \mathscr{C} < 1 - (1+\gamma)^{3/2} \sqrt{\frac{\alpha}{\gamma + \alpha} (1 + \frac{3^{4/3}\varepsilon^{2/3} \gamma }{(1+\gamma)^2})}
    + \gamma \sqrt{\alpha} + \mathcal{O}(\varepsilon)
.\end{equation}

\subsubsection{Additional priors: Parameterisations} \label{sec:code_additional_priors_parameterisations}

Additional priors can become necessary when using changes of variables. For example, in Hou's parameterisation for eccentricity (as in \hyperref[eq:hou_cv_full]{Eqs. \ref{eq:hou_cv_full}}), the value ranges for $e_c, e_s$ are individually defined as [-1, 1], but the sum of their squares cannot go beyond unity, since it would imply eccentricities higher than 1. Another prior enforcing the same eccentricity physical constraint becomes necessary.

\subsection{Stage 2:\ Compiling the run} \label{sec:write_script}

Python's process-based parallelism forces every worker to import identical code. To avoid repeated pickling of large data objects, \emp~auto-generates a secondary script that first loads the data and then defines model functions in situ. This structure removes pickling overhead and delivers near-linear scaling with dataset size.
Only the blocks actually requested are written to the script; unused packages and variables are omitted to keep the run time lightweight. Parallelisation back-ends are selected via a single string argument.

\subsection{Stage 3:\ Post-processing}\label{sec:code_postprocessing}

\paragraph{Posterior estimation:} For each parameter, several point estimates are provided: 1) the mean with 1, 2, and 3 standard deviations $\sigma_{\theta}$ as uncertainties; 2) the median with equivalent $n\sigma_{\theta}$ percentiles; and 3) the maximum of the posterior with equivalent high-density intervals (HDI).

Optionally, if model-comparison criterion is met, chains can be thinned and fed to a kernel density estimator (KDE) or a Bayesian Gaussian mixture (BGM) to obtain smoother distributions that can be used as priors for subsequent runs. This can be useful for diagnostic runs for signal searching, and should not be used for model comparison.

\vspace{-\baselineskip}
\paragraph{Statistical diagnostics:} 
\emp~records acceptance fractions, auto-correlation times, and computes evidence \lnz~and goodness-of-fit metrics: $\chi^2$ and $\chi^2_{\nu}$, measuring how well the model's predictions align with the observed data. The root-mean-square error (RMSE) is quanitified globally and for individual
instruments via RMSE and RMS$_i$, respectively. In addition, AIC, BIC, DIC, and HQIC were used for the model comparison. \emp~has the option of increasing the number of samples produced, until a pre-determined stopping criterion is met, such as the total length of the chain be equal to $n$ times the auto-correlation time, or by having the \lnz~value stable within a range for $n$ steps. This methodology enables a fully automated algorithm.

\vspace{-\baselineskip}
\paragraph{Model comparison and outputs:}
A tolerance-based driver based on the aforementioned goodness-of-fit metrics decides whether to add more signals and repeat the loop (default is $\Delta\mathrm{BIC}\geq5$). 
Comprehensive publication-ready figures are generated by an independent suite, which is coupled with parallel plotting and data aggregation techniques  to create them in minimal time.

\section{Benchmarks}  \label{sec:benchmarks}

Three benchmarks assessed \emp's ability to explore the posterior, conduct model comparison, and estimate evidence. The first benchmark uses real data from the well-known star 51 Pegasi, allowing direct comparison with published results to evaluate the sampler's performance. The second benchmark involves model comparison in a system with stellar activity, HD 55693. The final benchmark explores modelling nuances Barnard's star. These benchmarks will gauge both the performance and efficiency of the \emp~code.

\subsection{51 Pegasi}\label{sec:benchmarks_51Peg}

51 Pegasi (51 Peg) was the first exoplanet discovered using the RV method \citep{1995Natur.378..355M}. A quiet Sun-like star with low levels of stellar activity, it presents a single Keplerian signal with high S/N, indicating the nature of a hot Jupiter. With enough data and reasonable sampling, the signal is straightforward to detect, making it an ideal system to validate \emp, compare the efficiency of different samplers within the code, as well as comparing against a different exoplanet fitting tool, \juliet~\citep{2019MNRAS.490.2262E} in this case.

\vspace{-\baselineskip}
\paragraph{Stellar parameters and data:} 51 Peg is classified as a G2IV star with a visual magnitude of $V$=$5.46$, an optical colour of $B$-$V$=$0.67$, and an activity index of $\log{R'}{\mathrm{HK}}$=$-5.054$ \citep{2010ApJ...725..875I}. For this test, 256 LICK RV measurements were utilised, obtained from the Exoplanet Archive \footnote{http://exoplanetarchive.ipac.caltech.edu}, based on the work of \citet{2006ApJ...646..505B}.

\vspace{-\baselineskip}
\paragraph{Benchmark discussion:}
We first tested the null hypothesis $0K$: a white-noise model comprising offset, jitter, and a linear trend. Then we tested a model that adds a single Keplerian component ($1K$; \hyperref[eq:keplerian]{see Eq. \ref{eq:keplerian}}).

This benchmark is conducted in three modes, \emp~coupled with the \reddemc~sampler (emp-APT), \emp~with \dynesty~(emp-DNS), and \juliet~with \dynesty~(jul-DNS). The first comparison, emp-DNS with jul-DNS, seeks to validate \emp~as a fitting tool, as well as assess its performance. The second one, emp-APT with emp-DNS, validates the APT sampler \lnz~estimation, as well as its performance compared to DNS. Sampling efficiency is summarised by the effective samples per second, \texttt{enits} $\equiv \mathrm{ESS} / \mathrm{time}$, which reflects both the sampling quality and computational cost. Each configuration was repeated 11 times, we reported the mean and standard deviation across repeats.

Since no analytic evidence was available, we assessed the uncertainty calibration of the Bayesian evidence with a $\chi^2$ consistency test across repeats. For run $i$ with an estimate $\ln{\mathcal{\hat{Z}}}_i$ and a reported uncertainty of $\hat{\sigma}^2_{\mathcal{Z}, i}$, we compute
\begin{equation}
    \chi^2_{\sigma} = \frac{1}{N-1}\sum_{i=1}^{N}\left( \frac{\ln{\mathcal{\hat{Z}}}_i - \langle\ln{\mathcal{\hat{Z}}}\rangle}{\hat{\sigma}_{\mathcal{Z}, i}}\right).
\end{equation}

For $\chi^2_{\sigma}$, values $\gg$1 indicate uncertainty underestimation, $\ll$1 overestimation, and $\approx$1 well-calibrated uncertainty estimates.

After performing the emp-DNS runs (1500 live-points, stopping criterion $\Delta\ln{\hat{\mathcal{Z}}}<0.01$), \reddemc\,was set to work with (16, 256, 2048) for the number of temperatures, walkers, and steps, respectively, to roughly match the total run time of the  $K1$ of the other method. The Keplerian parameterisation that was available for \juliet\,was used, ($P$, $K$, $T_0$, $e$, $\omega$), with a period \Uniform{1}{100} and $T_0$\Uniform{0}{10}. Eccentricity and jitter used normal priors, \Normal{0}{0.3}, and \Normal{0}{15}, respectively. All other parameters used uniform priors in accordance to \hyperref[sec:code_default_bayesian_priors]{\S\ref{sec:code_default_bayesian_priors}}.

Performance differences are apparent: jul-DNS and emp-DNS (see \hyperref[tab:51peg_stats]{Table \ref{tab:51peg_stats}}), using the same sampler, have a similar sampling efficiency (around 3.75\% for the $0K$ and around 1.56\% for the $1K$). The wall time, for approximately the same number of evaluations, was considerably lower within \emp, at 21\% and 10\% of the total time, for the $0K$ and $1K$, respectively, leaving the enits in the same proportion.

Within \emp, the sampler differences between APT and DNS are evident as well. The wall time for APT is setup dependant, and since the same setup was used, the $0K$ and $1K$ runs have virtually the same wall-time. Therefore, the enits are used for comparing performance: for the $0K$, the enits are more than tripled with APT (2215.3 vs 666.3), while for the $1K$ they are almost doubled (594.4 vs 319.4).
For the APT, the evidence uncertainty $\chi^2_{\sigma}$ is slightly better calibrated in the $0K$ (1.19 vs 1.28) and much more conservative in the $1K$ (3.13 vs $\sim$12).

Evidence estimation is consistent between jul-DNS and emp-DNS, for the $0K$ model, they arrived at the same value $\ln{\hat{\mathcal{Z}}}=-1318.7\pm0.1;$  for the $1K$ model, emp-DNS and jul-DNS had $\ln{\hat{\mathcal{Z}}}=-884.12\pm2.07$ and $-884.53\pm1.96$, respectively.

On the other hand, within \emp, the APT and DNS are consistent only for the $0K$ model, with both sharing the estimate $\ln{\hat{\mathcal{Z}}}=-1318.7\pm0.2$; whereas, for the slightly more complex $1K$ model, APT presents a lower evidence, with a 2-$\sigma$ difference with respect to the DNS: $-880.64\pm0.52$ compared to $-884.12\pm2.07$. To solve this discrepancy, we incrementally increased the `chain-length' for both methods. In APT, the number of steps were increased up to 10\,000, but the estimates remained fairly similar, possibly flagging this $\ln{\hat{\mathcal{Z}}}$ as a stable estimate.

Similarly, increasing the DNS live-points up to 10\,000, showed no changes in the evidence. Nonetheless, by also reducing the stopping criterion to $\Delta\ln{\hat{\mathcal{Z}}}<0.001$, the estimate changed to $\ln{\hat{\mathcal{Z}}}=-880.82\pm2.58$ (for 11 runs as well), with a $\chi^2_{\sigma}=47.187$. This stricter estimation is consistent with the APT result. For these longer runs, the average run time was $1473.6\pm104.1$~s, while the efficiency was slightly increased to $1.74\pm0.02$, which resulted in $346.3\pm9.9$ enits.

When the phase-space has a complicated shape, there are subtleties that are hard to characterise.  For example, as eccentricity tends to zero, both angular parameters become increasingly degenerate. In this system, with eccentricity very close to zero, this degeneracy becomes hard to disentangle. For the longer DNS runs, which had a lower eccentricity estimation $e=0.006$ (see \hyperref[tab:51peg_params]{Table \ref{tab:51peg_params}}), we can see this phenomena here by looking at the angular parameters, their posteriors flatten, rendering higher uncertainties.

The best solution (see \hyperref[fig:51peg-model-redd]{Fig. \ref{fig:51peg-model-redd}} and \hyperref[fig:51peg-corner]{Fig. \ref{fig:51peg-corner}}) confirms a linear trend in the system, placing tight constraints on the trend and indicating the presence of a very distant massive companion.  The likely companion producing the trend has been imaged by astrometric surveys \citep{2011AJ....142..175R, 2019A&A...623A..72K} and so, the combination of these two detection methods could yield constrained companion properties (e.g. mass) once at least one RV inflection has been measured.

\begin{table}
    \caption{\label{tab:51peg_stats}51 Peg performance and evidence estimation.}
    \centering
    \begin{tabular}{llll}
        \toprule
        Stat        & emp-APT  & emp-DNS    & jul-DNS  \\
        \midrule
        \midrule
        0K \\
        \midrule
        time (s)    &220.5±5.7   &49.1±1.4      &234.1±1.6    \\
        eff (\%)    &11.37±0.16  &3.76±0.01     &3.73±0.01  \\
        enits       &\textbf{2215.3±74.5} &666.3±16.6    &139.5±0.4    \\
        $\ln{\hat{\mathcal{Z}}}$    &-1318.7±0.2 &-1318.7±0.1 & -1318.7±0.1  \\
        $\hat{\sigma}_{\mathcal{Z}}$&0.142±0.005   &0.076±0.001   & 0.076±0.001    \\
        $\chi^2_{\sigma}$           &\textbf{1.185}         &1.282         &1.289           \\
        \midrule
        1K \\
        \midrule
        time (s)    &218.8±3.22   &250.9±19.2   &2630.6±250.3     \\
        eff (\%)    &3.03±0.10    &1.65±0.04    &1.46±0.16        \\
        enits       &$\mathbf{594.4\pm23.3}$   &319.4±18.3   &30.4±2.4         \\
        $\ln{\hat{\mathcal{Z}}}$    &-880.64±0.52   &-884.12±2.07   &-884.53±1.96   \\
        $\hat{\sigma}_{\mathcal{Z}}$&0.192±0.027    &0.175±0.004    &0.175±0.004    \\
        $\chi^2_{\sigma}$           &$\mathbf{3.129} $     &12.530         &11.615         \\
        \midrule
        \bottomrule
    \end{tabular}
    \tablefoot{From left to right: \emp~with~\reddemc, \emp~with \dynesty, and \texttt{juliet} with \dynesty. From top to bottom:  white noise model ($0K$) and single Keplerian ($1K$), each tab contains descending the total run time in seconds, eff (the sampling efficiency), enits (effective samples per second), the Bayesian evidence estimate, its estimated uncertainty, and the $\chi^2$ test between internal and empirical evidence uncertainties.}
\end{table}

\begin{figure}
    \includegraphics[width=\columnwidth]{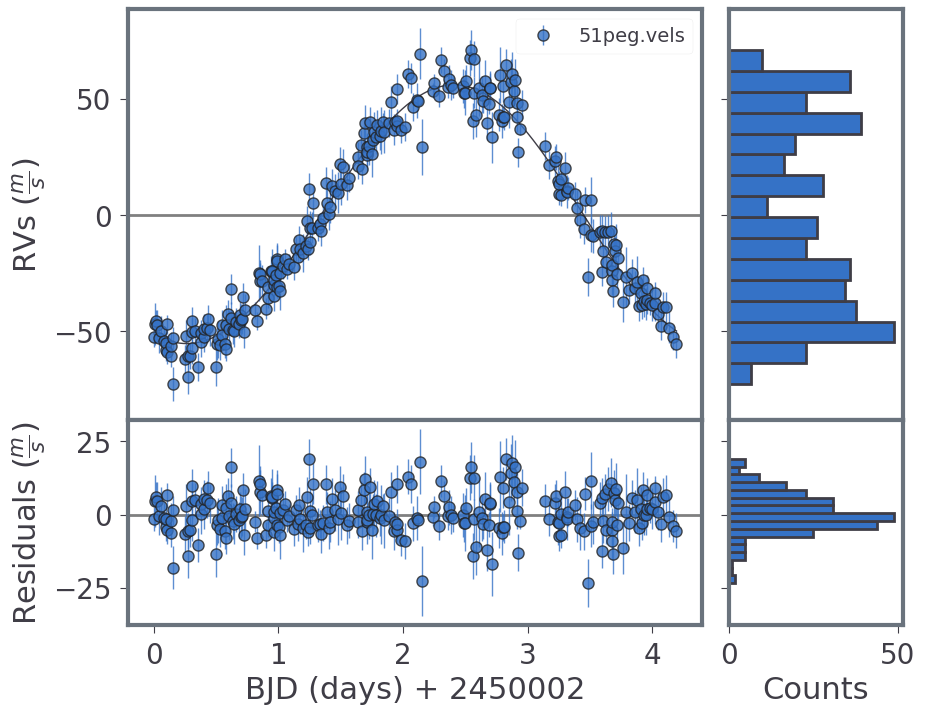}
    \caption{51 Peg Keplerian best fit model with \reddemc. Top: LICK RVs phase-folded to the period with the best-fit model (black line). Bottom: Residuals. Right: Histograms of the observations and residuals.}
    \label{fig:51peg-model-redd}
\end{figure}

\begin{figure} 
    \includegraphics[width=\columnwidth]{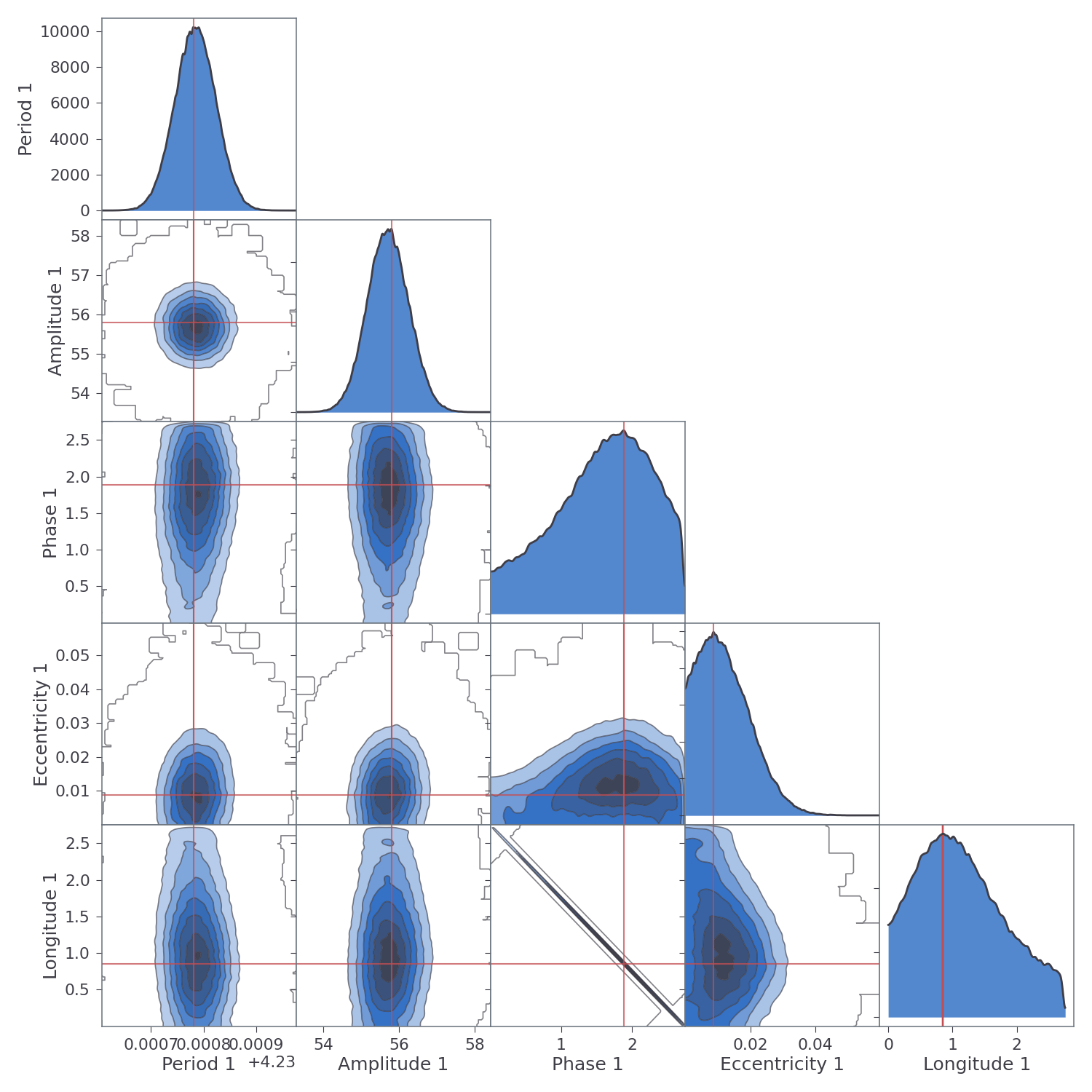}
    \caption{51 Peg corner plot of Keplerian parameters. Period, semi-amplitude, and eccentricity are well-defined narrow Gaussians. The angular parameters, $M_0$ and $\omega$, appear as wide Gaussians, highly correlated with each other, which is the case for $e=0$.}
    \label{fig:51peg-corner}
\end{figure}

\subsection{HD 55693}\label{sec:benchmarks_hd55693}

Disentangling stellar activity SA signals from those of exoplanets presents a significant challenge in RV studies \citep{2011IAUS..276..530D, 2016PASP..128f6001F}. Stellar activity can induce RV variations that resemble signals from orbiting exoplanets or interfere with their detection. These variations arise from stellar phenomena such as starspots, plages, and flares, which alter the line profiles of stellar spectra and, consequently, the measured RVs. Additionally, the timescales of stellar activity often overlap with orbital periods of exoplanets \citep{2001A&A...379..279Q, 2011IAUS..273..281B, 2018AJ....155..126D}, complicating the separation of these signals. Accurately characterising the host star's activity is therefore critical for confirming the presence of an exoplanet and to determine its properties with high precision.

The primary challenge lies in developing robust models capable of effectively distinguishing between stellar activity signatures and the subtle gravitational tug of an exoplanet. Such models often require advanced statistical approaches and long-term observational campaigns, including methods such as MA or GPR \citep{2014MNRAS.443.2517H, 2015MNRAS.452.2269R, 2019MNRAS.490.5585J, 2020A&A...639A..77S, 2023MNRAS.518.2627V}.

\vspace{-\baselineskip}
\paragraph{Stellar parameters and data:}
HD 55693 (HIP 34879) is classified as a G1.5V star \citep{2006AJ....132..161G} with a visual magnitude of $V=7.83$ and an optical colour of $B-V=0.67$. An activity index of $\log{R'}_{\mathrm{HK}}=-4.963$ was reported by \cite{2011A&A...528A.112L}, indicating a low activity level typical of a G-type star. These authors also identified a significant magnetic cycle with period of $P_{\mathrm{cyc}}$=$2403^{+266}_{-218}$~d and a stellar rotation period of $P_{\mathrm{rot}}$=27.4$\pm$3.2~d.

Two different spectrographs were used for the high-precision RVs--HARPS \citep{2000SPIE.4008..582P} and PFS \citep{2010SPIE.7735E..53C}. Since HARPS underwent an optical fibre upgrade in 2015, its RVs were divided into two datasets (before and after the upgrade). The observations were processed with the TERRA pipeline \citep{2012ApJS..200...15A}, resulting in two datasets designated TERRA1 and TERRA2, containing 29 and 19 RVs, respectively. Each RV is accompanied by stellar activity indices for the $S$-Index, full width at half maximum (FWHM), and bisector span (BIS). The PFS dataset consists of 36 RV measurements and corresponding $S$-Index values.

\vspace{-\baselineskip}
\paragraph{Benchmark discussion:}
This benchmark will showcase several noise models within the \emp~framework, as well as the model selection problem. From just the RV data, we  tried to recover the star's known rotation period and magnetic cycle from an uninformed perspective.

Six different noise models are compared: 1) WN: white noise only; 2) SA:\ linear correlations with stellar activities; 3) MA:\ an exponentially weighted moving average model; 4) SAMA:\ a combination of SA and MA; 5) \gprot:\  GP with a rotation kernel; and 6) \gprotplussho:GP with rotation and simple harmonic oscillator terms. Each model includes $K_n$ sub-models, where $n$ denotes the number of Keplerian signals. Consistent priors were used across models for equivalent parameters (see \hyperref[tab:hd55693_params]{Table \ref{tab:hd55693_params}}). A summary of model selection statistics is presented in \hyperref[tab:hd55693_stats1]{Table \ref{tab:hd55693_stats1}}.

An initial inspection of the Lomb-Scargle periodogram (see \hyperref[fig:hd55693_periodogram_h1]{Fig. \ref{fig:hd55693_periodogram_h1}} for TERRA1) reveals a strong period at 2403~d for RVs (matching the magnetic cycle identified by \cite{2011A&A...528A.112L}). The $S$-Index peaks at 2554~d, while the BIS shows a peak in the same range (though below the 10\% false-alarm-probability FAP threshold). A signal matching the known rotation period also appears in the RVs (at 29.7~d) and in the S-indices (26.1~d).

\begin{figure} 
    \includegraphics[width=\columnwidth]{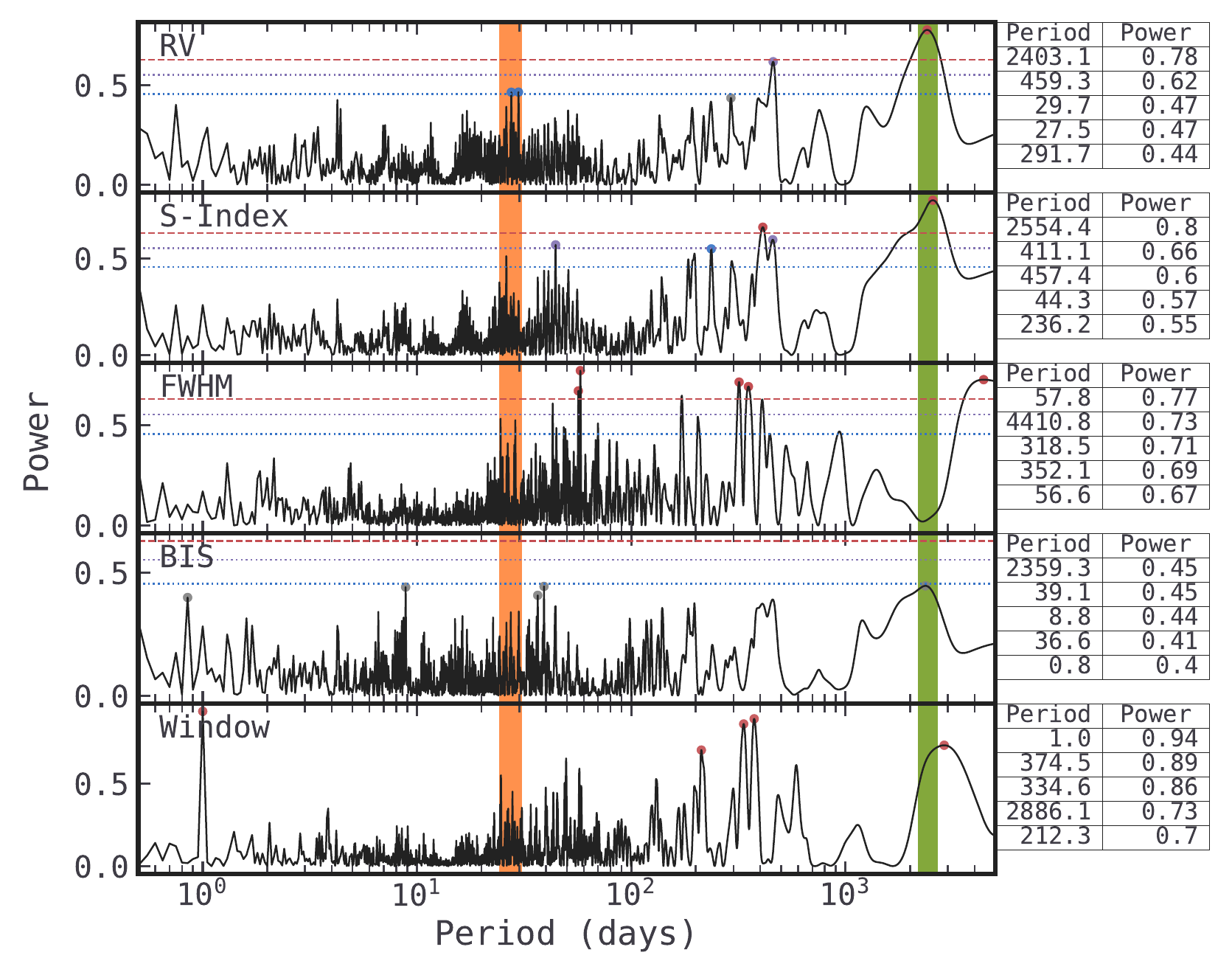}
    \caption{HD 55693 periodogram for TERRA1 data. Descending, RVs, $S$-Index, FWHM, BIS, and window function. FAP lines included for 10\%, 1\%, and 0.1\%, in dashed red, dotted purple, and dotted blue, respectively. Circle markers show the five periods with the greatest power, coloured by FAP region. The orange coloured region corresponds to $P_{\mathrm{rot}}=27.4\pm3.2$ and the green one to $P_{\mathrm{mag}}=2403^{266}_{-218}$.}
    \label{fig:hd55693_periodogram_h1}
\end{figure}

The correlogram for TERRA1 (see \hyperref[fig:hd55693_correlogram_h1]{Fig. \ref{fig:hd55693_correlogram_h1}}) shows a significant correlation between RVs and the $S$-index ($\rho$=0.80), as well as between RVs and BIS ($\rho$=0.56). Notably, the $S$-index and BIS are highly correlated at $\rho$=0.76, suggesting they might be tracing the same physical phenomena. Since only $S$-index is available in the PFS dataset, this index alone is employed as part of the SA modelling, for consistency across datasets.

In this preliminary analysis, the high correlation of $\rho$=0.80 implies contamination of the RVs by stellar activity and the LSP suggests that the $\sim$2500~d peak, shared between RVs and $S$-index is linked to stellar activity.
Other peaks that repeat in both measurements are at $\sim$459~d, $\sim$235~d, $\sim$26~d, and $\sim$193~d.

\begin{figure} 
    \includegraphics[width=\columnwidth]{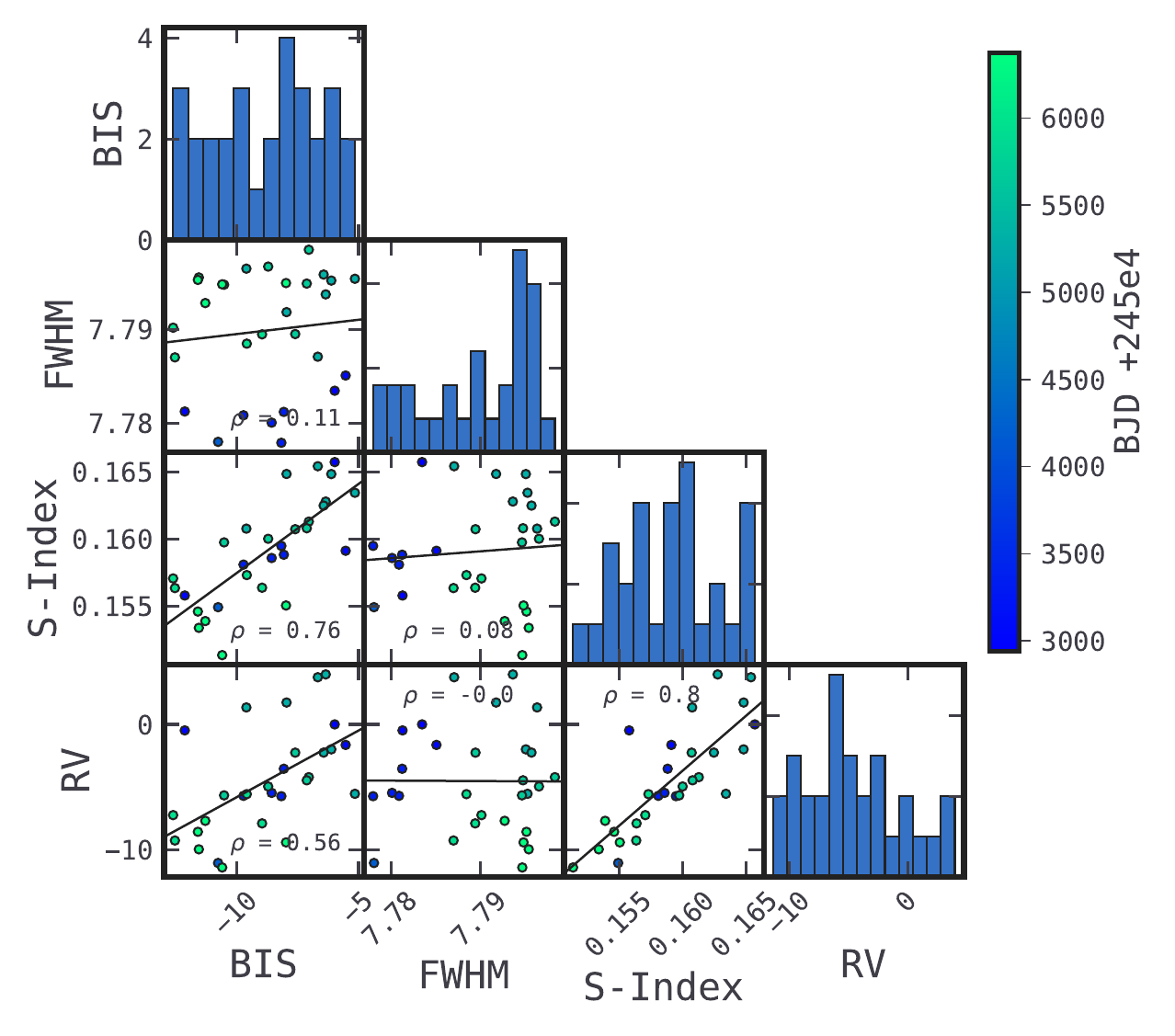}
    \caption{HD 55693 correlogram for TERRA1 data. Shows the Pearson Correlation coefficient ($\rho$) between RVs, $S$-Index, FWHM, and BIS. Diagonal displays the samples distribution. RV presents significant correlation with $S$-Index $\rho$=0.8 and BIS $\rho$=0.56. $S$-Index with BIS have $\rho$=0.76, suggesting they describe the same physical phenomena.}
    \label{fig:hd55693_correlogram_h1}
\end{figure}

For a star of this stellar-type and age, we expect a rotation period at $\sim$27-32 days \citep{2008ApJ...687.1264M}, and a magnetic cycle period at $\sim$2200-3300 days \citep{2011A&A...528A.112L}. This matches the signals appearing in both periodograms at $\sim$2500~d and $\sim$26~d. 
When using \emp~for the benchmarked models, each one finds first a significant detection at $\sim$2500~d, which we correspond with stellar activity, specifically, the magnetic cycle.

The periodogram of the RV residuals shows peaks under 10\% FAP at $P$=29.7, 56.0, and 19.8~d. And for $S$-index values below 0.1\% at $P$=396.3, 5529.8 and below 1\% at $P$=185.8, 36.7, 27.5, and 19.5~d. When adding a second signal, the WN($K_2$) and MA($K_2$) models found $P_2$=167~d. The \gprot($K_1$) and \gprotplussho($K_0$) models found $P_2$=$\sim$2.4~d. In addition, the SAMA($K_2$) model found $P_2$=$\sim$29.7~d. \reddtext{The multi-modal nature of the WN($K_2$) solution is illustrated in \reffig{fig:hd55693_post_per2}, where the posterior places comparable weight at $\sim$19, 21, 115 and 170~d. This explains both the broad HDI and the rejection of this solution.}

A priori, to select the best overall model, consistency across multiple metrics is a good starting point. \gprot($K_1$) has the highest evidence \lnz=-211.76 and the lowest RMS=0.81. Nevertheless, its reduced chi-square is well below unity (0.4), flagging over-fitting. For that reason, this model, as well as its cousin \gprotplussho($K_0$) are discounted. SAMA($K_2$) stands out: it has the second highest evidence \lnz=-212.03 amongst all models. Its $\chi^2_{\nu}$ indicates neither an over-, nor an under-fitting and its low RMS points to tight residuals.
The BIC is just 2.4 higher than the absolute minimum--WN($K_2$). This modest penalty is out-weighted by the much larger Bayes factor advantage ($e^{1.26}$$\approx$9.6).

Looking at how the evidence of each noise model evolves when adding signals can give some insight: For the WN model, $K_1$ presents very strong Bayes factor over the $K_0$, of $\approx$$e^{16.33}$. The $K_2$ over the $K_1$, although strong, $\Delta$\lnz=4.66, barely misses the strict $\Delta$\lnz=5 usually required in exoplanet detection.

The MA model should model the lower frequency modulation produced by the rotation period and exhibits a similar behaviour to the WN model, but with decreased effects, a much better $K_1$ over $K_0$ solution ($\Delta$\lnz=14.82). Furthermore, we see a worse improvement in the $K_2$ over the $K_1$, $\Delta$\lnz=0.21. This already gives some insight that this second signal might be a by-product of stellar rotation.

The SA model, finds the same $P_1$$\sim$2500~d, but it does not find a stable $P_2$, with different solutions (e.g. 86, 138, 167, 5200~d). This model should de-trend signals produced by stellar activity and it is noticeable by the fact that compared to the WN model, the evidence of improvement after adding this first signal is much lower $\Delta$\lnz=4.30 (and therefore, rejected) compared to the WN's 16.33. The correlation coefficients also go down considerably (e.g. from $\mathcal{A}_{\text{T1}}$=0.81 to 0.44). The SAMA model, including both effects, shows an improvement of $\Delta$\lnz=2.36 when adding the first signal, the lowest so far. This is a moderate indicator that these signals are produced by stellar activity.

With all of this in mind, a conservative conclusion to this RV analysis would be that there are no planets in this system, a magnetic cycle with a period of $\sim$2500~d, and a high frequency rotation period not well characterised within the RV data.  Furthermore, since the true \pmag~and \prot~are known, models describing these phenomena are expected to best describe the system, either by removing the stellar activity noise--like the SA($K_0$), SAMA($K_0$), or \gprotplussho($K_0$) models, or by modelling the activity as signals--like the WN($K_2$), MA($K_2$), or \gprot($K_1$) models.

The former approach, removing the stellar activity noise, can be seen naturally in this benchmark, by the selection of either the SA($K_0$) or SAMA($K_0$) as best model. The simplicity of SA($K_0$) and its capacity to remove competing signals from the dataset, makes it an attractive plausible physical description of the system. 

A simple prior on the SA coefficients support this logic: we compare the SA($K_0$) model with different priors for the stellar activity coefficient $\mathcal{A}_{INS}$: 1) default \Uniform{-1}{1};  2) constrained around the correlation coefficient $\rho$ and its error \citep{pearson-error} $\frac{(1-\rho^2)}{\sqrt{N-3}}$; and 3) with a Normal prior $\mathcal{N}(\rho, \sigma_{\rho})$. The values for $\rho$ and $\sigma_{\rho}$ can be found in \hyperref[tab:hd55693_SA_stats]{Table \ref{tab:hd55693_SA_stats}}, as well as the metric values. The SA solutions present the same parameter solutions, reflected in the same RMSE=2.73.
Notably, these two models, unbounded-normal and bounded-uniform, have similar evidences, but present an improvement of 4.06 and 4.59 over the \Uniform{-1}{1} SA($K_0$) model. An additional signal under these priors gives \lnz=-212.20±0.08 and -212.54±0.13, respectively.

On the other hand, the latter approach, modelling the noise as signals, would require additional information to arrive at, but would give out important details about the system; namely, the \prot~ and \pmag~of the star. The SAMA($K_2$) model features the best overall evidence for non-overfitting models, characterising \prot=$29.7^{+0.01}_{-0.02}$~d (see \hyperref[fig:hd55693_prot]{Fig. \ref{fig:hd55693_prot}}) and \pmag=$2557^{+70}_{-38}$~d (see \hyperref[fig:hd55693_pmag]{Fig. \ref{fig:hd55693_pmag}}), in accordance to the presumed true periods for this system. Additional work on the system would be required, such as fitting signals on the $S$-index data alone to impose stronger priors in the RVs or a simultaneous fit with RVs and $S$-index, enforcing shared parameters between signals. This is further explored in the following benchmark.

\begin{figure} 
    \includegraphics[width=\columnwidth]{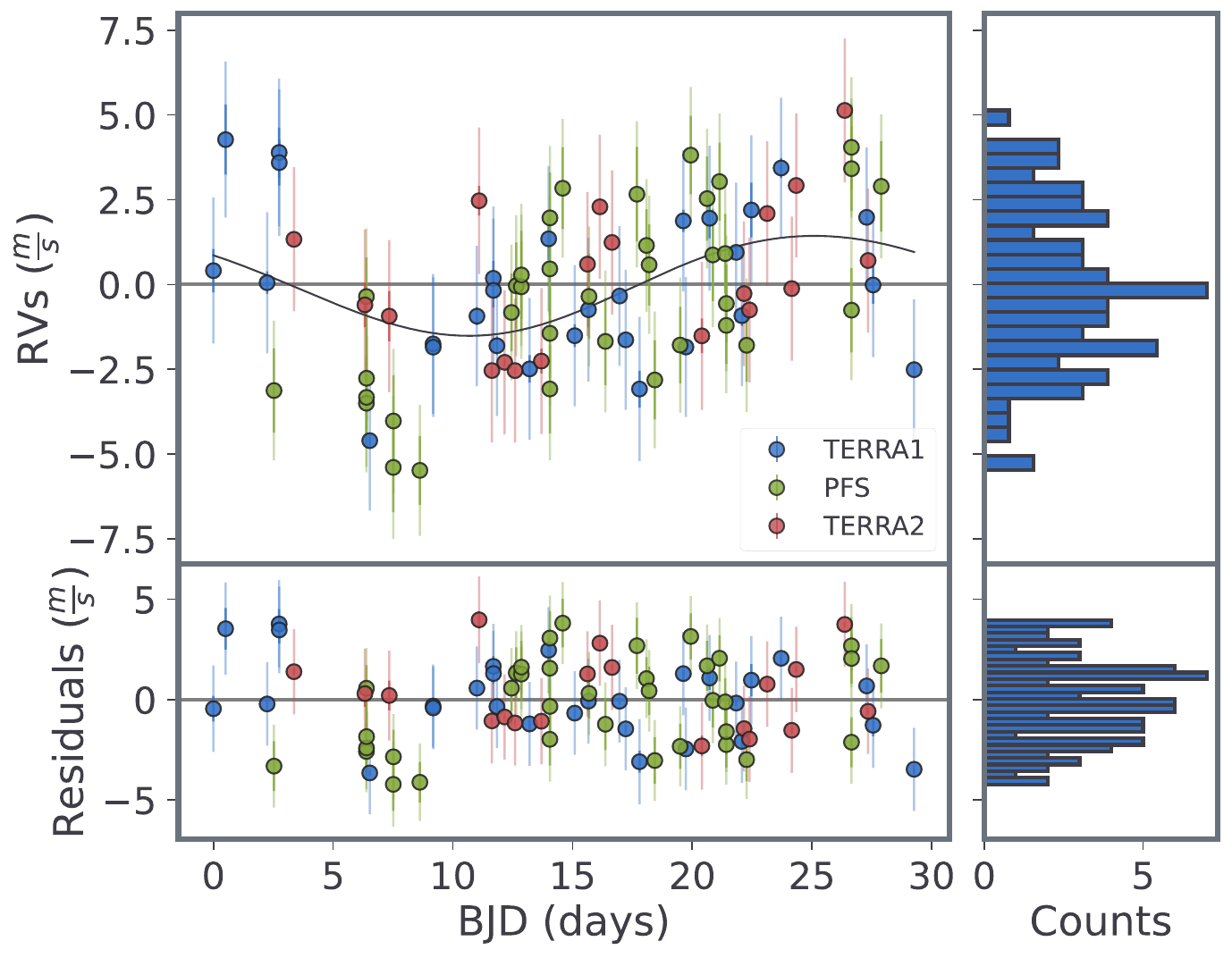}
    \caption{\label{fig:hd55693_prot}HD 55693 SAMA(K2) model. Top: RVs phase-folded to \prot=$29.7^{+37.2}_{-15.3}$~d with the best-fit model (black line). Marker colours differentiate datasets, blue (T1), green (PFS), and red (T2). Bottom: RV residuals. Right: histograms of observations and residuals.}
\end{figure}

\begin{figure} 
    \includegraphics[width=\columnwidth]{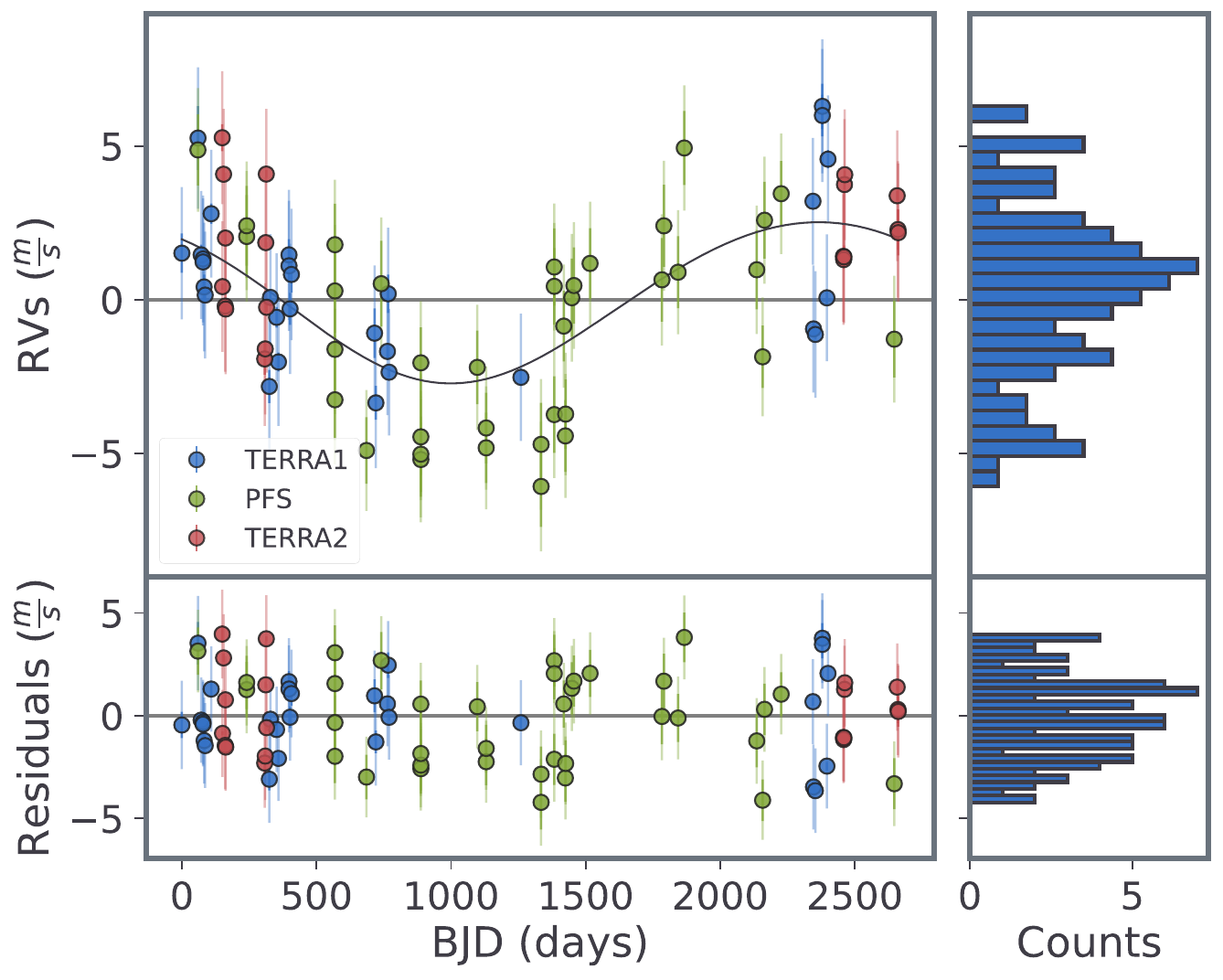}
    \caption{\label{fig:hd55693_pmag}HD 55693 SAMA(K2) model. Top: RVs phase-folded to \pmag=$2531.6^{+76.9}_{-8.0}$~d with the best-fit model (black line). Marker colours differentiate datasets, blue (T1), green (PFS), and red (T2). Bottom: RV residuals. Right: Histograms of the observations and residuals.}
\end{figure}

\begin{figure} 
    \includegraphics[width=\columnwidth]{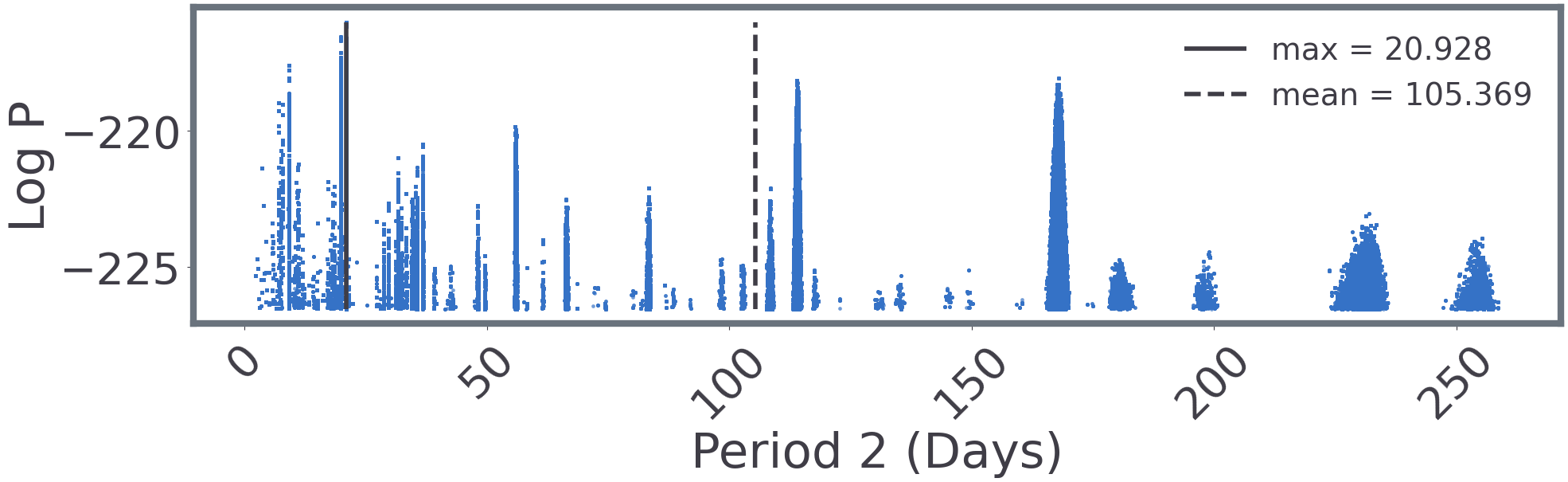}
    \caption{\reddtext{HD 55693 WN(${K_2}$) model, top of the posterior for \prot.}}
    \label{fig:hd55693_post_per2}
\end{figure}

\subsection{Barnard's star}\label{sec:benchmarks_GJ699}

M-dwarf systems, being smaller than the Sun, possess their habitable zone closer to their host. This characteristic makes them prime candidates in the search for Earth-like planets \citep{2007AsBio...7...85S, 2008A&A...485..571J, 2016PhR...663....1S}. Barnard's star is not only an M-dwarf but also the closest single-star system to the Sun, at a distance of about six light years. It exhibits the highest known proper motion and a very low level of stellar activity, making it an excellent target in the search for Earth analogues.

\vspace{-\baselineskip}
\paragraph{Stellar parameters and data:} \label{sec:gj699-stellar-params-data}

Barnard's star (GJ 699) is classified as an M3.5V-M4V star with a visual magnitude of $V$=$9.51$ and an optical colour of B$-$V=$-$1.73 \citep{2015A&A...577A.128A}. \citet{2019MNRAS.488.5145T} reported a low activity level, $\log{R'}_{\mathrm{HK}}$=$-$5.82, a rotation period of $P_{\mathrm{rot}}$=145$\pm$15~d, and a magnetic cycle of $P_{\mathrm{mag}}$=3800$\pm$600~d. \citet{2024A&A...690A..79G}, hereafter GH24, based on 156 ESPRESSO RVs over four years found $P_{\mathrm{rot}}$=$136.2^{+10.5}_{-9.4}$~d and $P_{\mathrm{mag}}$=$3325^{+276}_{-226}$~d, along a Keplerian signal at $P_{\mathrm{Kep}}$=3.1533$\pm$0.0006~d.

We follow GH24 and analyse 792 RVs obtained with Carmenes (CAR), HARPS (H15), HARPS-N (HAN), and ESPRESSO. Owing to the ESPRESSO fibre-link upgrade, its data is split into pre- and post-upgrade subsets (E18 and E19). After discarding ESPRESSO observations with uncertainties >0.5\ms, and H15/HAN points with RV uncertainties >0.85\ms\,or FWHM uncertainties >2.5\ms, the final sample comprises 479 CAR, 100 H15, 48 HAN, 9 E18, and 140 E19 measurements (N=776).

\vspace{-\baselineskip}
\paragraph{Benchmark discussion:}

The FWHM data were examined first to identify any trends or correlations. Multiple models were tested for the RV data, both with and without incorporating information from the FWHM analysis. This procedure was carried out in parallel for the stand-alone ESPRESSO dataset and for the combined ESPRESSO+HAN+H15 datasets \reddtext{(LSP in \reffig{fig:gj699-periodogram-all})}.

\begin{figure} 
    \includegraphics[width=\columnwidth]{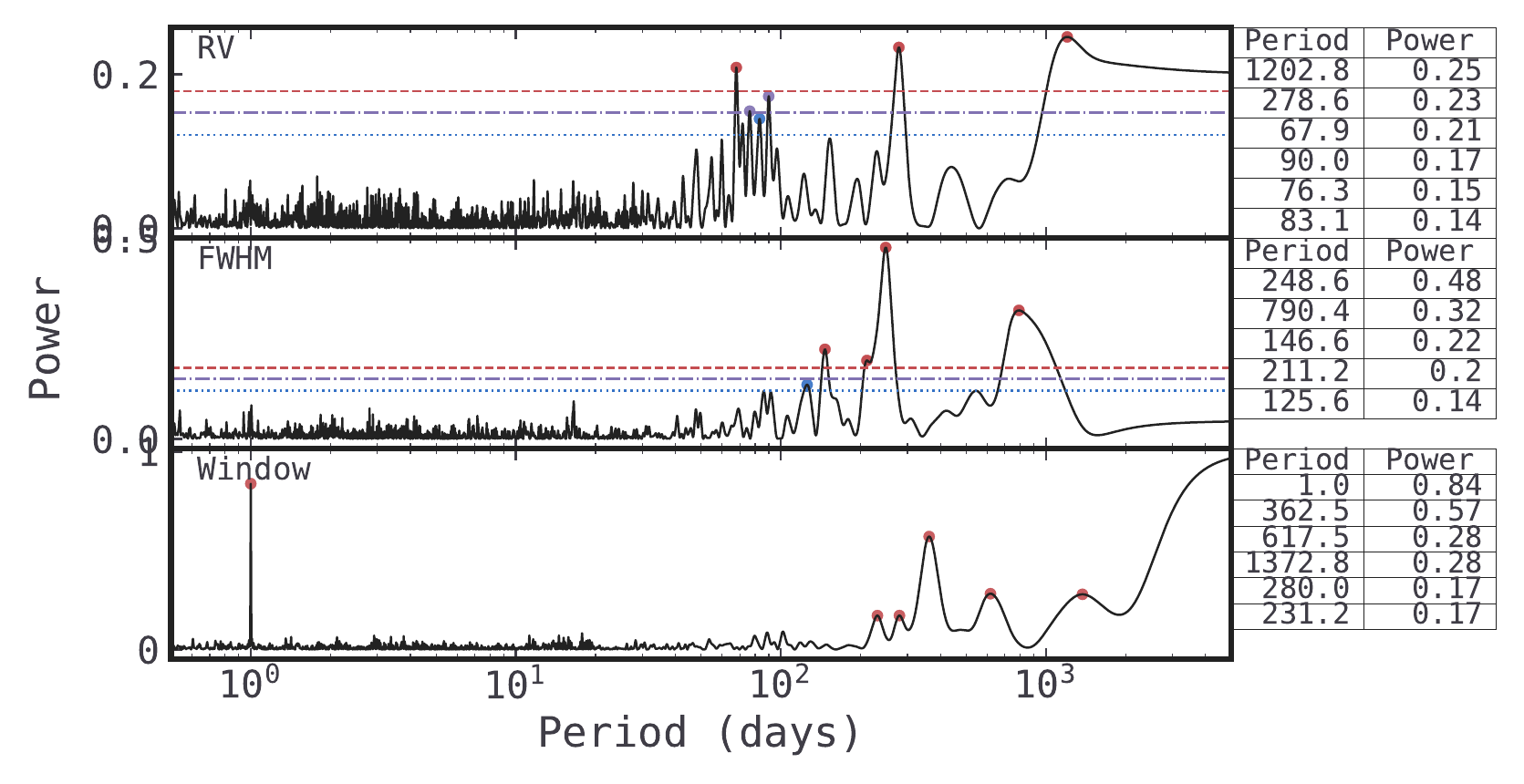}
    \caption{Barnard's star periodogram for ESPRESSO data. Descending, RVs, FWHM, and window function. FAP lines included for 10\%, 1\% and 0.1\%, in dashed red, dash-dotted purple, and dotted blue, respectively. Circle markers show the five periods with the greatest power, coloured by FAP region.}
    \label{fig:gj699-periodogram-espresso}
\end{figure}

The LSP for ESPRESSO FWHM data (see \hyperref[fig:gj699-periodogram-espresso]{Fig. \ref{fig:gj699-periodogram-espresso}}) revealed significant peaks at 249, 790, 147, and 211~d, along with a smaller peak at 126~d. The 249~d peak could correspond to the one-year alias of the 147~d signal, whilst the small bump at 211~d (and the one at 126~d) may be caused by differential rotation. A correlation of $\rho$=0.43 was found between RV and FWHM (see \hyperref[fig:gj699_correlogram]{Fig. \ref{fig:gj699_correlogram}}), suggesting stellar activity contamination in the RVs.

\emp~was first applied to the ESPRESSO FWHM data with uninformative wide priors (see WN results in \hyperref[tab:gj699_stats_fwhm]{Table \ref{tab:gj699_stats_fwhm}}). The white-noise-only (WN) approach finds first a 245~d signal, followed by 166~d, and 2384~d, each with increasing probability. The first two signals are of special interest, since they also appear in the LSP, and can be recognised as the one-year alias of the rotation period and $P_{\text{rot}}$  from \citet{2019MNRAS.488.5145T}.

\begin{figure} 
    \includegraphics[width=\columnwidth]{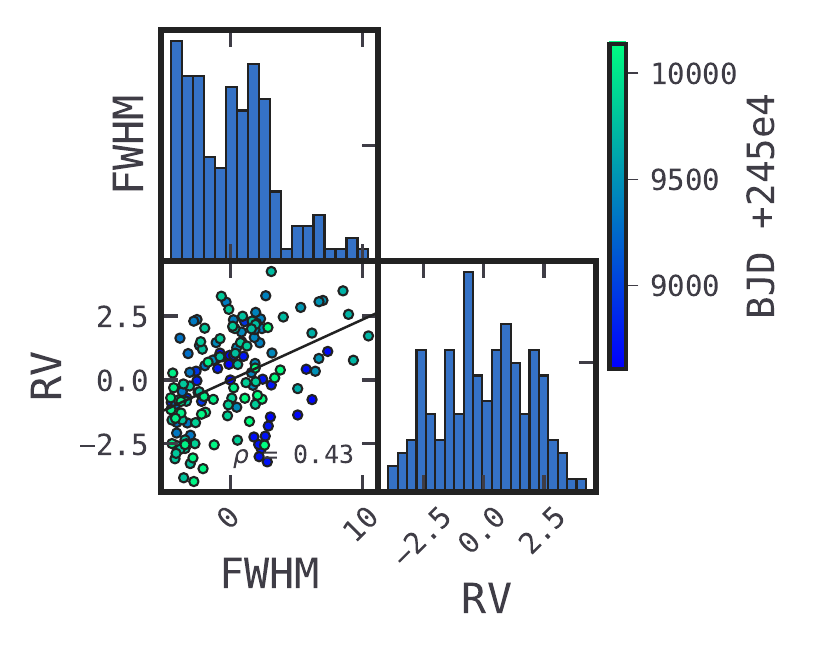}
    \caption{Barnard's star correlogram for E18 and E19 data. Displays the Pearson correlation coefficient ($\rho$) between RVs and FWHM. Diagonal displays the samples distribution. The RV presents significant correlation with FWHM $\rho=0.43$.}
    \label{fig:gj699_correlogram}
\end{figure}

The rotation period was then modelled with GPs. To minimise over-fitting, some sensible priors were applied: the rotation period was constrained to \Uniform{50}{300}, covering both the presumed rotation period and its one-year alias, and a normal prior centred on the RMS of the data of \Normal{3.3}{3.3} was imposed on its amplitude to prevent the GP from absorbing all observed variability. The most revealing GP models used a rotation kernel (\gprot, defined in \hyperref[eq:gprot]{Eq. \ref{eq:gprot}}), which resulted in a fit $P_{\text{rot}}$ of 168 d (see \hyperref[tab:gj699_params_fwhm]{Table \ref{tab:gj699_params_fwhm}} for parameters comparison). Adding a single Keplerian models the 250~d period, whereas adding a magnetic cycle model instead (defined in \hyperref[eq:magnetic-cycle]{Eq. \ref{eq:magnetic-cycle}}) models a $P_{\text{mag}}$ of 506~d, whilst reducing $P_{\text{rot}}$ to 155~d. A tighter prior on the long cycle period, \Uniform{800}{5000}, produced a $P_{\text{mag}}$ of 809~d. Inspecting the posterior shows a boundary solution at a truncated peak, likely the previous 506~d solution. An even tighter prior of \Uniform{2000}{5000} resulted in $P_{\text{rot}}$$\sim 250$~d and a flat posterior for most of the magnetic cycle parameters, including $P_{\text{mag}}$. None of these variants of \prot+Mag conveyed a clear peak at $\sim 3200$ days.

The \gprot($K_0$) model has an evidence equal to $-$215.86±0.01 and the inclusion of an additional Keplerian ($K_1$) provides a marginally worse evidence ($\Delta$\lnz$=$$-$0.34); so \gprot($K_0$) is selected as the best model. The proposed value of $P_{\mathrm{rot}}$ is detected as \prot=$170.95^{+22.01}_{-1.98}$~d, but the long-period cycle was not detected in any model featuring uniform priors, likely due the comparatively short baseline of the ESPRESSO data (1533~d).

A GP model variation with \gpgrot~(see \hyperref[eq:gp-grot]{Eq. \ref{eq:gp-grot}}), similar to the approach of GH24, was also tested. Although it found \prot=$174.2$, it was ruled out based on evidence comparison. The inclusion of HARPS and HAN data presents consistent results with those of ESPRESSO on its own (see \hyperref[tab:gj699_stats_fwhm_eh]{Table \ref{tab:gj699_stats_fwhm_eh}}), presenting the best \lnz~ and BIC values.

Next up, the RV data were analysed, informed by the previous FWHM results. The base model includes two signals: one for stellar rotation and one for the magnetic cycle. For stellar rotation, the period boundaries were mildly constrained to \Uniform{50}{300} once again, while for the magnetic cycle, elusive to the previous FWHM analysis, the priors from \cite{2019MNRAS.488.5145T} were adopted, \Normal{3250}{300}, matching the priors used by GH24.

Amongst the tested models, we included \gpthreesho: two SHO kernels for the rotation and one for the magnetic cycle; \gpgrot$+S$:\ a rotation kernel (see \hyperref[eq:gprot]{Eq. \ref{eq:gprot}}) and a simple sinusoid to describe the magnetic cycle (see \hyperref[eq:sinusoid-model]{Eq. \ref{eq:sinusoid-model}}); \gprotplussho:\  a rotation kernel and a SHO kernel to describe the magnetic cycle);  $\mathcal{GP_{\mathrm{2rot}}}$: two rotation kernels, one for each period; and \gpgrot$+$Mag:\ a variation in the rotation kernel (described in \hyperref[eq:gp-grot]{Eq. \ref{eq:gp-grot}}) and a double sinusoid (\hyperref[eq:magnetic-cycle]{Eq. \ref{eq:magnetic-cycle}}), akin to the model used in GH24. After the initial evaluation, \emp~added a Keplerian with a period of \Uniform{0.5}{50}.

All models found signals consistent with $P_{\mathrm{rot}}$=140~d, $P_{\mathrm{mag}}$=3200~d and $P_{\mathrm{Kep}}$=3.15~d (see \hyperref[tab:gj699_params]{Table \ref{tab:gj699_params}}). Based on the statistics (see \hyperref[tab:gj699_stats_rv]{Table \ref{tab:gj699_stats_rv}}), \gprot$+S$, \gprot+Mag, \gpgrot$+S$, and \gpgrot+Mag had similar statistics and parameters, and \gpgrot$+S$ is selected as the best model. This model returned a marginally higher evidence over \gpgrot+Mag $\Delta Z$=0.69 and the reduced complexity of using two fewer parameters argues in favour of this model. In addition, the $\Delta (\ln P$-\lnz) is of 3.15 between the two models, remarking on the uniqueness of the \gpgrot$+S$ solution. The phase-folded model for the Keplerian is illustrated in \hyperref[fig:gj699-model-espresso]{Fig. \ref{fig:gj699-model-espresso}}.

A re-run of this model was performed, with a \Uniform{800}{5000} prior instead of \Normal{3250}{300} for \pmag, to test its dependence on the normal prior. The resulting parameters were \prot=$140.67^{+7.60}_{-9.61}$~d, \pmag=$2486^{+401}_{-19}$~d, and $P_{\mathrm{Kep}}$=$3.15^{+0.98}_{-1.70}~d$. Although this magnetic cycle is shorter than the $\sim$3200~d presumed signal, its upper bound is poorly constrained by the limited temporal baseline. Crucially, the Keplerian signal remains at 3.15~d. These results align with GH24 (see \hyperref[tab:gj699_planet_comp]{Table \ref{tab:gj699_planet_comp}}).

\begin{table}
   \caption{\label{tab:gj699_planet_comp}Planet parameters of GJ 699 b.}
   \centering
   \begin{tabular}{lll}
   \toprule
   Parameter                  & This work               & GH24                  \\
   \midrule
   \midrule
   $P_{\mathrm{Kep}}$ (days)  & 3.1536$\pm$ 0.0003  & 3.1533 $\pm$ 0.0006   \\
   $K_p$ (ms$^{-1}$)          & 0.56 $\pm$ 0.03     & 0.55 $\pm$ 0.07       \\
   $m_{p}$ sin$I$ (\me)       & 0.38 $\pm$ 0.03     & 0.37 $\pm$ 0.05       \\
   $a_p$ (AU)                 & 0.02315 $\pm$0.00039& 0.02294 $\pm$ 0.00033 \\
   $e_p$                      & <0.04               & <0.16                 \\
   \bottomrule
   \end{tabular}
\end{table}

\begin{figure} 
    \includegraphics[width=\columnwidth]{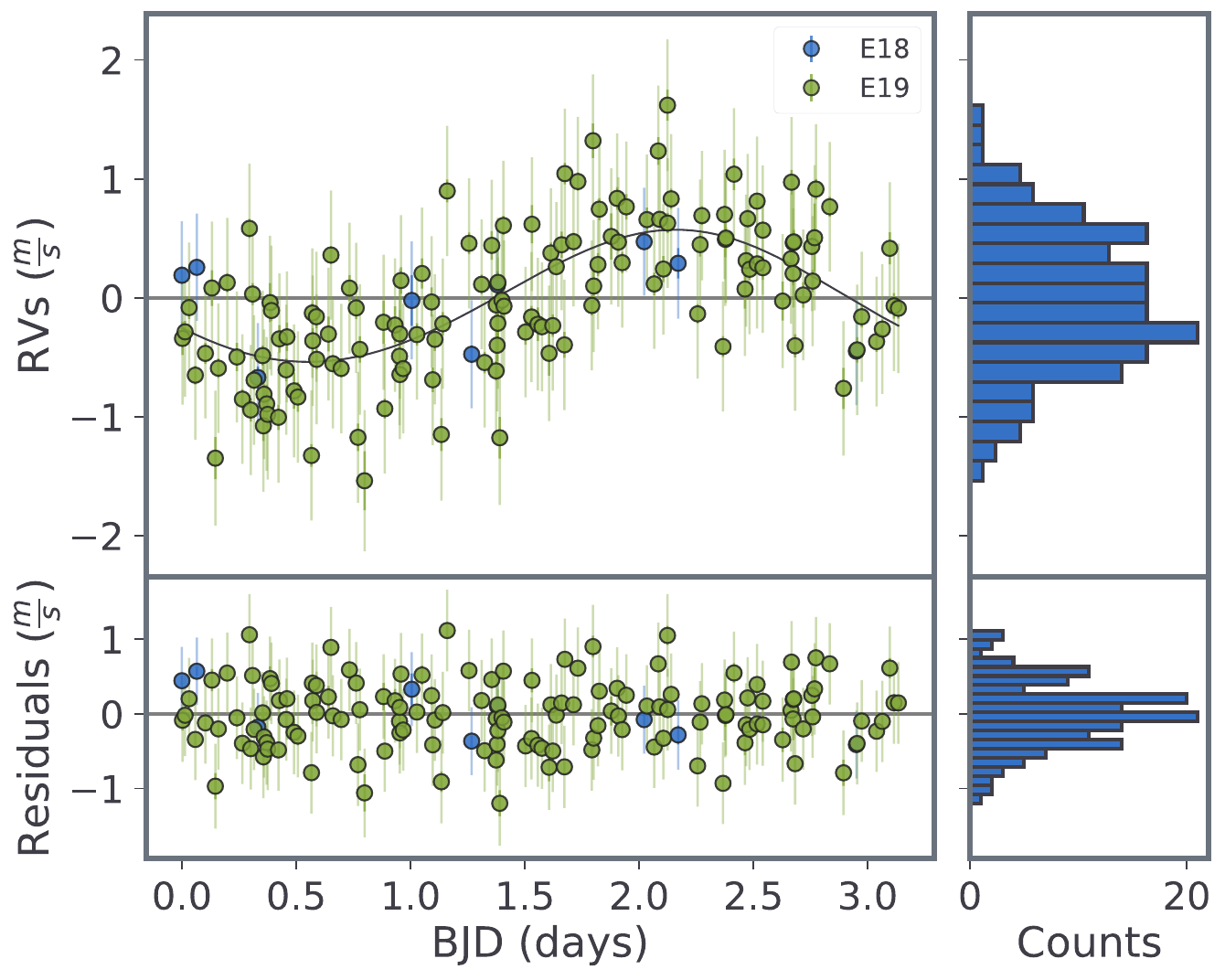}
    \caption{\label{fig:gj699-model-espresso}Barnard's star \gpgrot$+S$ model. Top: RVs phase-folded to $P=3.1533$~d with the best-fit model (black line). Marker colours differentiate datasets, blue (E18), and green (E19). Bottom: RV residuals. Right: Histograms of observations and residuals.}
\end{figure}

\begin{figure} 
    \includegraphics[width=\columnwidth]{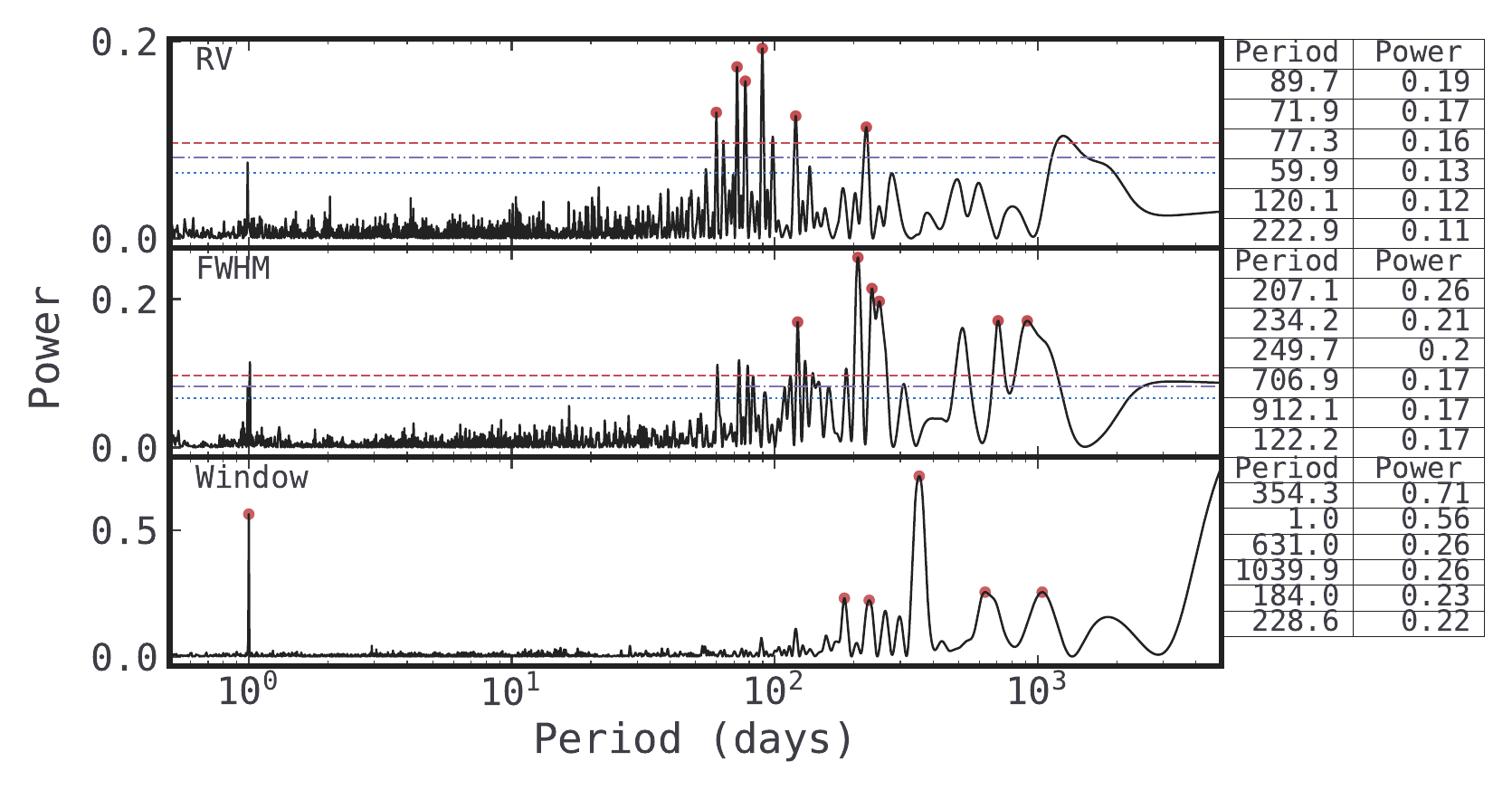}
    \caption{Barnard's star periodogram for the combined ESPRESSO, HARPS, and HARPS-N data. Descending, RVs, FWHM, and the window function. FAP lines included for 10, 1, and 0.1\%, in dashed red, dash-dotted purple, and dotted blue, respectively. Circle markers show the 10 periods with the greatest power, coloured by FAP region.}
    \label{fig:gj699-periodogram-all}
\end{figure}

\section{Discussion} \label{sec:discussion}

Bayesian inference has become a fundamental tool in exoplanet detection to push the detectability limits in RV data. Several challenges arise: the highly multi-modal nature of the parameter space, correlations between parameters, data sparsity, low S/N, contamination due to stellar activity, and other complicating factors. \emp~addresses these issues by employing an original APT MCMC sampler and \reddemc, as well as an ample selection of noise models. As shown in the benchmarks with 51 Peg (see \hyperref[sec:benchmarks_51Peg]{\S\ref{sec:benchmarks_51Peg}}), the increase in effective computation speed due to the sampling algorithm is almost two-fold compared to \dynesty's DNS. And the increase due to the \emp~environment is at least four-fold, compared to a different RV fitter software using the same sampler. This directly translates to a plethora of models getting tested in an equivalent time with other tools. \emp~also provides the necessary apparatus to make model building (and model comparison) easy and straightforward.

While nested sampling excels in estimating Bayesian evidence (\lnz), the APT approach excels in thoroughly exploring the parameter space, which (in the presence of complex noise structures) allows the algorithm to retrieve precise exoplanetary parameters. This trade-off highlights the importance of selecting the appropriate method based on the specific goals of an analysis, whether it is to prioritise robust evidence estimation or refine parameter constraints. It is important to bear in mind that dynamic nested sampling tends to underestimate the evidence's error, while thermodynamic integration tends to overestimate it. In the case of 51 Peg, \emp~successfully retrieved planetary parameters with high precision, APT outperforming NS (see \hyperref[tab:51peg_params]{Table \ref{tab:51peg_params}}).

Statistical methods alone are insufficient for a robust exoplanet detection, as they struggle to distinguish between planetary and stellar activity. Effective model selection requires the integration of domain knowledge, heuristics, and multiple statistical indicators instead of just one. Furthermore, an over-reliance on the Bayesian evidence might hinder the unveiling of the system. As observed in the HD 55693 analysis (see \hyperref[sec:benchmarks_hd55693]{\S\ref{sec:benchmarks_hd55693}}), a completely uninformed approach to the system could lead to the presence of one or two planets (the WN $K_2$ model presenting a higher evidence than its predecessors, WN $K_0$ and WN $K_1$). By introducing stellar activity analysis, the first signal (P$\sim$2500~d) becomes suspect of being activity-related and not Keplerian at all. The same applies for secondary signals at P$\sim$259, 235, 193, and 26~d. By applying some domain knowledge on stars for this spectral type and age \citep{2008ApJ...687.1264M, 2011A&A...528A.112L}, it becomes reasonable to expect a magnetic cycle period between 2200-3300~d and a rotation period around $\sim$30~d.

By trying to model the activity as both rotation and a magnetic cycle, but without normal priors to guide our sampling, with different noise models, several interesting points arise: 1) the MA model effectively removed the rotation signal, fitting exclusively the magnetic cycle, whilst reducing its amplitude compared to other solutions; 2) the SA model managed to remove the rotation as well, but as a single Keplerian was added, the correlation coefficient parameter dropped enough to make room for this signal. Adding a prior on the coefficient severely reduces the degeneracy induced by the Keplerian, effectively removing both the magnetic cycle and the rotation; 3) with a GP for the rotation, a solution with the best evidence is obtained. Nevertheless, the rotation period it fits $P_{rot}$=2.45~d is much lower than expected. A deeper look at its statistics reveals an overly low $\chi^2_{\nu}$=0.41, hinting towards an over-fitting; therefore this model should be treated with caution or discarded.

This benchmark underscores the necessity of cautious prior selection and comparative model evaluation. A rigid reliance on evidence estimation alone may lead to misleading conclusions when prior assumptions are incomplete or introduce biases.

While GPR is a powerful tool for mitigating stellar noise, its application requires careful tuning. As demonstrated in the Barnard's star analysis, five different GP models arrived at the same solution for \prot, \pmag, and $P_{\mathrm{Kep}}$, with similar \lnz~values between them. Almost as a cautionary tale, the model with the best \lnz~was the one that required the least number of parameters (\gpgrot+S).

\section{Conclusions}  \label{sec:conclusions}

Three systems were analysed with our new \emp~code--51 Peg, HD 55693, and Barnard's star (GJ 699). For 51 Peg \refsec{sec:benchmarks_51Peg}, using 256 LICK RVs we were able to reproduce the results of \citet{2006ApJ...646..505B} with tighter posteriors on the parameters  for 51 Peg b: $P$=$4.230782^{+0.000016}_{-0.000013}$~d, $K$=$55.69^{+0.18}_{-0.23}$~m/s, and $e$$<$0.01. The \emp~code outperforms another RV fitting tool (using the same sampler) by a factor of at least 4, and the \reddemc~APT sampler is shown to outperform by almost a factor of 2 over \dynesty's DNS sampler.

For HD 55693, three datasets were used, TERRA1, TERRA2, and PFS, with 29, 19, and 36 observations, respectively. The data were subjected to preliminary LSP and correlogram analysis, which revealed a strong correlation between $S$-index and RVs ($\rho=0.80$), as well as matching periods for a presumed \pmag~and \prot. With uninformed priors on the periods, the best model was shown to be a GP with a rotation kernel proves to be false, leading to an unphysically short rotation period. The SA model, composed of stellar activity linear correlations, works best with its prior informed (even modestly) by the correlations, leaving no Keplerian signals in the fit. The SAMA model, a mixture of SA with an exponentially weighted moving average found the periods, \pmag=$2557^{+70}_{-38}$ and \prot=$29.72^{+0.01}_{-0.02}$, consistent with the presumed true values. Although with just RV data and uninformative priors, this last model does not have enough evidence to supersede the non-Keplerian model. The importance of applying domain knowledge and heuristics to the model's priors. In addition,  the dangers of over-relying on a single metric for model comparison were also demonstrated by performing small modifications on the priors' boundaries.

For Barnard's star, the work by \citep{2024A&A...690A..79G} was followed, to test \emp\, on recent discoveries that make use of advanced noise modelling as well as high precision RVs on low-S/N data. The presumed \prot~value was found on both the LSP and the \reddemc~run of the FWHM data, but \pmag~did not appear on any of them without applying a strong prior based on a $\mathrm{CaHK}$ analysis done previously by \citet{2019MNRAS.488.5145T}. In the analysis of the RV data, the planet signal at $P_{\mathrm{Kep}}=3.15$~d was consistently recovered across models. Moreover, removing the prior for this long-term cycle found a different value for \pmag=$2486^{+401}_{-19}$~d, but the Keplerian signal was still successfully retrieved $P_{\mathrm{Kep}}$=$3.15^{+0.98}_{-1.70}~d$. In this case, other peaks in the posterior were boosted enough to maintain them in the 95\% HDI range used for uncertainties, but still far away from the 3.15~d peak (e.g. compared to the second-highest peak, at 4.12~d, the posterior difference $\Delta p(\pmb{\theta} | D, M) \sim$ 7).

The results presented in this paper show that the \emp~code is an effective and flexible tool for planet detection and characterisation using RV data. Its key advantages are its ability to explore a broad multi-modal parameter space efficiently, as well as its modularity, which allows various models to be seamlessly integrated, and its computing performance, which democratises the process of exoplanet detection. The implementation of the APT framework significantly improves convergence rates and ensures that the global posterior maximum is consistently identified. Including GP and MA methods, as well as different depictions of low-frequency modulations, is crucial for mitigating stellar activity contamination. The evidence-based model selection approach employed by \emp~provides a systematic way to determine the most probable model, in the least amount of time.

Future developments of the \emp~code will focus on integrating additional data types, notably photometry and astrometry \citep[with the optimised Gaia+Hipparcos-based method developed by][]{2023MNRAS.525..607F}, to facilitate a unified and simultaneous analysis of multiple observational channels. Further enhancements to stellar activity modelling are also anticipated, ensuring more robust discrimination between planetary signals and stellar noise. In addition, by applying \emp~to extensive RV catalogues, it will be \reddtext{plausible} to conduct population-level studies that unveil broader trends in exoplanet demographics. Through these advancements, \emp~is poised to become a more comprehensive and adaptive platform for exoplanet detection and characterisation, thereby conquering new realms in the uncharted frontiers of planetary systems.

\begin{acknowledgements}
    All benchmarks were performed on a computer with an AMD Ryzen Threadripper 3990X 64-core processor with 128 GB of DDR4-3200Mhz RAM, limiting the use to 24 physical cores and 24 threads. This manuscript was written with Overleaf.
      We are grateful to Fabo Feng and Mikko Tuomi for stimulating early discussions on planet detection, and to Mar\'ia Maass and Ra\'ul Melo for feedback on the initial draft. PAPR and JSJ gratefully acknowledge support by FONDECYT grant 1240738 and from the ANID BASAL project FB210003. We also greatly appreciate the support from the CASSACA China-Chile Joint Research Fund through grant CCJRF2205. 
\end{acknowledgements}

\bibliographystyle{aa}
\bibliography{aa54336-25}

\begin{appendix}

\section{Available samplers}\label{sec:appendix_samplers}

\subsection{\texttt{reddemcee}}\label{sec:sampler_reddemcee}

This model is an APT MCMC based on the excellent \emc~and \texttt{ptemcee} \citep{2016MNRAS.455.1919V}.
Chain tempering has been shown to be necessary to efficiently sample highly multi-modal posteriors \citep{2007MNRAS.381.1607G,2013A&A...551A..79T}, where instead of sampling the posterior of the distribution, a modified posterior is sampled. The modification of the posterior is made to artificially dampen the strength of the posterior maxima, bringing it closer to the posterior noise floor. Normally, the inverse temperature $\beta \in [0,1]$ is used, such that the likelihood will look like $p(\pmb{\theta} | M)^{\beta}$.

The benefit from this implementation is that now the samplers at different temperatures can build proposal densities that are based on chains with other temperatures, and since the walkers in the hotter chains are less constrained, they are less likely to get stuck in regions of the posterior that are much higher than others, bringing confidence to the fact that cold chain ($\beta$=1) members have sampled the actual maximum of the posterior and have not gotten trapped in a region of high probability that is not the global maximum.

Another benefit from this method is that with multiple chains at different temperatures, one is able to approximate the Bayesian evidence, through thermodynamic integration \citep{2005PCCP....7.3910E, 2004AIPC..707...59G} or Stepping Stones \citep{steppingstones2011}. Given that \emp~is developed primarily to do a thorough search in a wide parameter space, the use of parallel tempering is preferred and implemented as the native format, in the form of \reddemc~\citep{reddemcee-paper}.

\subsection{\texttt{emcee}}\label{sec:sampler_emcee}

The MCMC Affine-Invariant Ensemble sampler \emc\footnote{http://dan.iel.fm/emcee/current/}
\citep{2013PASP..125..306F, 2012ApJ...745..198H}.

\emc~makes use of a `stretch move' to allow the algorithm's many $w$ walkers to sample the posterior independently, yet with the collective nature of the $w-1$
ensemble. This means that the proposal density of each walker is based on the current positions of the full $w-1$ walker set, and not just the position of the single previous walker, common to other sampling methods like Metropolis-Hastings or Gibbs samplers. Therefore, the algorithm can produce independent samples in much shorter auto-correlation times when compared to these other samplers. 

However, when testing \emc~on various data sets, two big issues were found when sampling. In a lot of cases extremely long chains were necessary in order for the samplers to converge to the posterior maximum, due to the high multi-modality. Also, walkers had a tendency to get stuck in local minima, depending on the shape of the posterior. Both of these issues are addressable by constraining the phase-space around the target posterior peak, which requires, of course, knowledge of the solution before-hand.

\subsection{\dynesty}\label{sec:sampler_dynesty}

Although the parallel tempering method is highly recommended for broad searches in multi-modal phase-spaces, \dynesty~is an alternative Bayesian posterior sampling engine which uses DNS, Dynamic Nested Sampling \citep{2019S&C....29..891H}, a generalisation of the standard nested sampling (SNS) algorithm where the live-points (akin to MCMC walkers) vary in number to improve sampling efficiency. In SNS, there is a fixed amount of live-points and most of the computational effort is spent iterating towards posterior peaks. Dynamically varying the points used per posterior peak allows the algorithm to shrink its prior faster and more thoroughly, and to arrive at more precise estimations (more details on SNS and DNS can be found in the aforementioned papers as well as \citealp{2004AIPC..735..395S}).

The tests performed have shown that when using Uniform sampling the run time is exceedingly high for wide priors, taking an exorbitant amount of time to converge. Constraining the phase-space is needed using this scheme. With that said, \dynesty~is excellent as either an alternative or complementary sampler to \texttt{reddemcee} for multi-modal posterior distributions, given that the prior volume is not unconstrained.

\subsection{\texttt{PyMC3}}\label{sec:sampler_pymc3}

PyMC3 is a well known and widely used MCMC Python package. It offers many sampling algorithms, but the crown jewel is the  NUTS code \citep{2015arXiv150708050S}. It is particularly useful for models with many continuous parameters, taking into account the posterior density gradient for steps, allowing it to meet convergence criteria extremely fast. PyMC3 in particular has several self-tuning strategies for adaptively exploring the posterior distribution, whilst being powered by \texttt{Theano} \citep{2016arXiv160502688T} to transcode to C for improved performance.

NUTS uses a scaling matrix method, which gives a rough shape of the distribution, meaning it does not take vastly different sized steps across dimensions. This matrix is based on the sample variance obtained during the tuning phase (analogous to the burn-in phase for MCMC). As such, efficiency is lost on too differently scaled parameter searches, as well as multi-modal or non-Gaussian distributions, and based on the empirical tests done, this method is not recommended for wide searches.

\newpage
\onecolumn
\section{51 Peg}\label{sec:appendix_51peg}

\begin{table*}[h!]
    \caption{\label{tab:51peg_params}51 Peg b model parameters.}
    \centering
    \begin{tabular}{lllll}
        \toprule
Parameter            & \reddemc  & dyn-long & dyn-short & Butler et al. 2006   \\
        \midrule
        \midrule
P (d)           &$4.230782^{+0.000016}_{-0.000013}$ &$4.230781^{+0.000038}_{-0.000038}$&$4.230784^{+0.000014}_{-0.000014}$ &$4.230785^{+0.000036}_{-0.000036}$ \\
K (\ms)         &$55.69^{+0.18}_{-0.23}$            &$55.66^{+0.54}_{-0.54}$           &$55.71^{+0.20}_{-0.19}$ &$55.94^{+0.69}_{-0.69}$            \\
$T_0$ (d)       &$1.55^{+0.16}_{-0.38}$             &$0.24^{+1.27}_{-1.35}$            &$1.56^{+0.23}_{-0.17}$ &$1.51^{+0.61}_{-0.61}$             \\
e               &$0.0085^{+0.0108}_{-0.0067}$       &$0.0059^{+0.0076}_{-0.0043}$      &$0.0102^{+0.0032}_{-0.0029}$ &$0.0130^{+0.0120}_{-0.0120}$       \\
$\omega$ (rad)  &$1.07^{+0.24}_{-0.55}$             &$4.51^{+1.46}_{-1.73}$            &$1.08^{+0.34}_{-0.26}$ &$1.01$                             \\
        \midrule
$\dot{\gamma}$ (\ms yr$^{-1}$) &$-1.62^{+0.09}_{-0.05}$&$-1.60^{+0.19}_{-0.19}$  &$-1.60^{+0.07}_{-0.07}$&$-1.64^{+0.16}_{-0.16}$ \\
$\gamma$ (\ms)                 &$5.41^{+0.08}_{-0.26}$ &$5.32^{+0.44}_{-0.44}$   &$5.31^{+0.16}_{-0.16}$ &-                       \\
$\sigma$ (\ms)                 &$0.64^{+0.01}_{-0.59}$ &$1.034^{+0.841}_{-0.709}$&$1.05^{+0.27}_{-0.27}$ &-                       \\
        \bottomrule
    \end{tabular}
    \tablefoot{From left to right:\ parameter estimates for \emp~with \reddemc~, \dynesty~random-slice (dyn-rs), and the Butler et al. (2006) solution. $T_0$ was substracted 2450000d for readability. In Butler 2006, $\omega$ had ill defined uncertainties, due the proximity of $e$ to 0. The parameters displayed here are from the run with the median maximum-likelihood for each method.}
\end{table*}

\newpage
\onecolumn
\section{HD 55693}\label{sec:appendix_hd55693}

\begin{table*}[h]
\caption{\label{tab:hd55693_stats1}HD 55693 model statistics.}
\centering
\begin{tabular}{lcccccc}
\toprule
 & $\ln{P}$-\lnz                & \lnz       & \lnzerr     & BIC & $\chi^2_\nu$ & RMS \\
\midrule
\midrule
WN ($K_0$)                      &11.48±0.01  &-235.28±0.01 & 0.25 &474.19±0.01 &1.09±0.01 &3.54±0.01 \\
\textbf{WN} ($K_1$)             &23.68±0.03  &-218.95±0.01 & 0.43 &439.29±0.06 &1.16±0.01 &2.54±0.01 \\
WN ($K_2$)                      &33.97±0.30  &-214.29±0.10 & 0.09 &431.53±0.45 &1.22±0.16 &2.11±0.01 \\
\midrule
MA ($K_0$)                      &13.84±0.10  &-227.94±0.11 &0.07 &463.64±0.01 &1.11±0.01 &3.15±0.01 \\
\textbf{MA ($K_1$)}             &22.13±0.09  &-213.12±0.05 &0.09 &439.57±0.16 &1.18±0.07 &2.41±0.02 \\
MA ($K_2$)                      &33.70±0.86  &-212.91±0.18 &0.10 &438.18±1.67 &1.36±0.01 &2.06±0.02 \\
\midrule
\textbf{SA ($K_0$)}             &3.48±0.06  &-217.79±0.06 &0.09 &438.89±0.07 &1.16±0.04 &2.73±0.01 \\
SA ($K_1$)                      &25.99±0.12 &-213.49±0.08 &0.08 &437.04±0.17 &1.23±0.05 &2.27±0.01 \\
SA ($K_2$)                      &35.54±1.51  &-213.29±0.04 &0.09 &439.69±3.01 &1.05±0.22 &1.99±0.08 \\
\midrule
\textbf{SAMA ($K_0$)}           &-1.23±0.22  &-215.74±0.20 &0.10 &440.54±0.28 &1.22±0.06 &2.60±0.01 \\
SAMA ($K_1$)                    &32.22±0.05  &-213.38±0.01 &0.09 &440.59±1.66 &1.27±0.02 &2.12±0.15 \\
SAMA ($K_2$)                    &41.57±0.28  &-212.03±0.07 &0.11 &433.97±0.54 &1.27±0.04 &1.85±0.01 \\
\midrule
\gprot ($K_0$)                  &28.01±0.01  &-235.17±0.01 &0.22 &463.06±0.02 &1.09±0.011 &2.68±0.01 \\
\textbf{\gprot ($K_1$)}         &31.10±0.59  &-211.76±0.12 &0.10 &432.21±1.16 &0.40±0.074 &0.81±0.05 \\
\midrule
\textbf{\gprotplussho ($K_0$)}  &27.55±0.13  &-214.70±0.12 &0.18 &436.34±0.13 &0.41±0.010 &0.81±0.04 \\
\bottomrule
\end{tabular}
\tablefoot{Runs from different models: ($K_n$) denotes the number of Keplerian signals. In descending order: white noise only (WN), exponentially weighted moving average (MA), stellar activity linearly correlated (SA), stellar-activity plus moving average (SAMA), Gaussian Process with a rotation kernel (\gprot), and GP with a rotation kernel plus a simple harmonic oscillator (\gprotplussho).}

\end{table*}

\begin{table*}[h]
    \caption{\label{tab:hd55693_SA_stats} HD 55693 WN vs SA model statistics.}
    \centering
    \begin{tabular}{lcrcl}
    \toprule
    & \multicolumn{1}{c}{WN} & \multicolumn{3}{c}{SA} \\
    Statistic & $K_1$    &$\mathcal{U}(-1, -1)$ & $\mathcal{U}$($\rho$-$\sigma_{\rho}$, $\rho$+$\sigma_{\rho}$) & $\mathcal{N}(\rho, \sigma_{\rho})$\\
    \midrule
    \midrule
    $\ln{P}$-\lnz      &23.68±0.03      &3.48±0.06    &4.75±0.03    &4.52±0.04    \\
    \lnz               &-218.95±0.01    &-217.79±0.06 &-213.20±0.01 &-213.73±0.01 \\
    BIC                &439.29±0.06     &438.89±0.07  &438.76±0.04  &438.83±0.05  \\
    \(\chi^2_\nu\)     &1.16±0.01       &1.16±0.04    &1.12±0.02    &1.14±0.03    \\
    RMSE               &2.54±0.01       &2.73±0.01    &2.73±0.01    &2.73±0.01    \\
    \bottomrule
    \end{tabular}
    
    \tablefoot{From left to right: WN($K_1$) model, SA($K_0$) model with different $\rho$ priors, \Uniform{-1}{1}, \Uniform{\rho-\sigma_{\rho}}{\rho+\sigma_{\rho}}, and \Normal{\rho}{\sigma_{\rho}}. Coefficient values are: $\rho_1$=$0.80$, $\sigma_{\rho1}$=$0.07$; $\rho_2$=$-0.01$, $\sigma_{\rho2}$=$0.20$; $\rho_3$=$0.56$, $\sigma_{\rho3}$=$0.13$.}
\end{table*}

\newpage

\begin{table*}[h!]
\caption{\label{tab:hd55693_params} HD 55693 model parameter estimations.}
\centering
\begin{tabular}{lllllllll}
\toprule
   Parameter & Prior &\textbf{WN ($K_1$)}  & WN ($K_2$)  & \textbf{MA ($K_1$)} & MA ($K_2$) & SA ($K_0$) & SAMA ($K_2$) & \gprot ($K_1$)\\
\midrule

\midrule
\multicolumn{9}{c}{Acceleration, offsets and jitter} \\
\midrule

$\gamma_{\text{T1}}$ (\ms) &$\mathcal{U}$(-8.5, 8.5) & $-1.17^{+0.5}_{-0.15}$& $-1.95^{+1.36}_{-0.68}$ &$-1.46^{+0.79}_{-0.7}$ &$-1.32^{+0.47}_{-0.22}$ &$-1.2^{+0.29}_{-0.16}$ &$-1.38^{+0.19}_{-0.42}$ &$-1.55^{+0.63}_{-0.64}$ \\
$\gamma_{\text{PFS}}$ (\ms)&$\mathcal{U}$(-6.7, 6.7) & $0.71^{+0.37}_{-0.14}$&  $1.21^{+0.56}_{-0.66}$ & $0.4^{+0.79}_{-0.61}$ & $1.39^{+0.59}_{-0.77}$ &$-0.12^{+0.28}_{-0.27}$& $1.26^{+0.58}_{-0.72}$ &  $0.75^{+0.31}_{-0.3}$ \\
$\gamma_{\text{T2}}$  (\ms)&$\mathcal{U}$(-5.7, 5.7) & $0.11^{+1.11}_{-1.44}$&  $0.14^{+1.17}_{-1.64}$ &$0.64^{+1.12}_{-2.06}$ & $0.98^{+1.31}_{-2.52}$ &$0.3^{+0.16}_{-0.27}$  &$-0.55^{+0.22}_{-0.62}$ & $0.23^{+1.19}_{-1.34}$ \\
$\sigma_{\text{T1}}$  (\ms)&$\mathcal{N}$(1, 5)      & $2.01^{+0.37}_{-0.41}$&  $2.42^{+0.42}_{-0.38}$ &$1.79^{+0.72}_{-0.41}$ & $2.15^{+0.35}_{-0.02}$ &$2.37^{+0.23}_{-0.1}$  & $1.63^{+0.52}_{-0.42}$ & $0.16^{+0.04}_{-0.16}$ \\
$\sigma_{\text{PFS}}$ (\ms)&$\mathcal{N}$(1, 5)      & $2.67^{+0.08}_{-0.34}$&  $1.56^{+0.96}_{-0.49}$ &$2.51^{+0.09}_{-0.27}$ & $1.27^{+1.06}_{-0.48}$ &$3.05^{+0.22}_{-0.18}$ & $1.53^{+0.79}_{-0.47}$ & $1.02^{+0.14}_{-0.43}$ \\
$\sigma_{\text{T2}}$  (\ms)&$\mathcal{N}$(1, 5)      & $2.32^{+0.52}_{-0.52}$&   $2.41^{+0.37}_{-0.1}$ &$2.02^{+0.52}_{-0.47}$ &  $1.92^{+0.62}_{-0.5}$ &$1.77^{+0.33}_{-0.39}$ & $1.99^{+0.27}_{-0.11}$ & $1.04^{+0.73}_{-0.02}$ \\

\midrule
\multicolumn{9}{c}{Magnetic cycle} \\
\midrule

$P_1$ (days)    &$\mathcal{U}$(1, 6712)   & $2463.4^{+143.8}_{-96.3}$    & $2523.6^{+86.7}_{-81.3}$    & $2395.5^{+188.5}_{-93.5}$    &$2540.1^{+43.6}_{-44.9}$  &- &$2557.0^{+70.1}_{-36.7}$   &$2457.7^{+95.3}_{-87.0}$    \\
$A_1$ (\ms)     &$\mathcal{U}$(1e-6, 7.1) &        $4.4^{+0.15}_{-0.38}$ &      $4.43^{+0.52}_{-0.56}$ &       $3.97^{+0.03}_{-0.19}$ &   $3.98^{+0.02}_{-0.24}$ &- &    $2.82^{+0.16}_{-0.15}$ &     $4.33^{+0.49}_{-0.55}$ \\
$M_{01}$(rad)   &$\mathcal{U}$(0, 2$\pi$) &       $0.79^{+0.39}_{-0.79}$ &      $2.85^{+1.76}_{-2.04}$ &          $0.4^{+1.4}_{-0.4}$ &   $3.93^{+2.36}_{-1.84}$ &- &    $2.05^{+1.54}_{-0.09}$ &      $0.47^{+1.48}_{-0.1}$ \\
$e_1$           &$\mathcal{N}$(0, 0.1)    &         $0.2^{+0.14}_{-0.2}$ &      $0.14^{+0.06}_{-0.14}$ &       $0.17^{+0.03}_{-0.17}$ &   $0.11^{+0.06}_{-0.11}$ &- &    $0.17^{+0.03}_{-0.12}$ &        $0.2^{+0.0}_{-0.2}$ \\
$\omega_1$ (rad)&$\mathcal{U}$(0, 2$\pi$) &       $5.73^{+0.54}_{-0.53}$ &      $3.77^{+2.46}_{-1.71}$ &        $5.9^{+0.38}_{-1.41}$ &   $2.86^{+3.38}_{-1.79}$ &- &    $4.75^{+0.11}_{-1.35}$ &      $6.05^{+0.23}_{-1.1}$ \\

\midrule
\multicolumn{9}{c}{Stellar rotation} \\
\midrule

$P_2$ (days)          &$\mathcal{U}$(1, 300)   &- & $167.23^{+83.81}_{-162.33}$ &- & $167.13^{+13.64}_{-59.25}$ &- &   $29.72^{+0.01}_{-0.02}$ &- \\
$A_2$ (\ms)           &$\mathcal{U}$(0, 7.1)   &- &      $2.95^{+1.07}_{-1.63}$ &- &      $4.31^{+1.37}_{-3.0}$ &- &    $3.03^{+1.52}_{-1.57}$ &- \\
$M_{02}$(rad)         &$\mathcal{U}$(0, 2$\pi$) &- &      $0.99^{+5.01}_{-0.99}$ &- &     $0.76^{+1.41}_{-0.66}$ &- &    $5.16^{+0.37}_{-0.39}$ &- \\
$e_2$                 &$\mathcal{N}$(0, 0.1)    &- &      $0.74^{+0.15}_{-0.13}$ &- &     $0.83^{+0.16}_{-0.07}$ &- &    $0.67^{+0.29}_{-0.26}$ &- \\
$\omega_2$ (rad)      &$\mathcal{U}$(0, 2$\pi$) &- &      $3.29^{+1.38}_{-0.32}$ &- &     $3.74^{+0.74}_{-0.97}$ &- &     $2.52^{+0.2}_{-0.81}$ &- \\

\midrule 
\gprot$\rho$ (days)   &$\mathcal{U}$(1, 300)    &- &- &- &- &- &- & $2.45^{+0.04}_{-0.26}$ \\
\gprot$\sigma$ (\ms)  &$\mathcal{U}$(0, 7.1)    &- &- &- &- &- &- & $1.16^{+1.83}_{-0.25}$ \\
\gprot$f$             &$\mathcal{U}$(0, 10)     &- &- &- &- &- &- & $0.01^{+0.99}_{-0.01}$ \\

\midrule
\multicolumn{9}{c}{Noise} \\
\midrule
MA $\phi$       &$\mathcal{U}$(-1, 1)  &- &- & $0.58^{+0.16}_{-0.23}$ &   $0.41^{+0.05}_{-0.1}$ &- &    $0.1^{+0.2}_{-0.17}$ &- \\
MA $\tau$ (days)&$\mathcal{U}$(1, 300) &- &- &$7.68^{+16.18}_{-4.91}$ &$27.88^{+7.65}_{-25.81}$ &- &$56.33^{+2.06}_{-54.83}$ &- \\

\midrule
$\mathcal{A}_{\text{T1}}$ &$\mathcal{U}$(-1, 1) &- &- &- &- &$0.81^{+0.07}_{-0.04}$  &$0.41^{+0.05}_{-0.06}$ &- \\
$\mathcal{A}_{\text{PFS}}$&$\mathcal{U}$(-1, 1) &- &- &- &- &$-0.09^{+0.06}_{-0.06}$ &$0.11^{+0.11}_{-0.18}$ &- \\
$\mathcal{A}_{\text{T2}}$ &$\mathcal{U}$(-1, 1) &- &- &- &- &$0.58^{+0.04}_{-0.07}$  &$0.48^{+0.16}_{-0.15}$ &- \\

\bottomrule
\end{tabular}
\tablefoot{Models with $K_n$ Keplerians in parenthesis, from left to right: White noise (WN), exponentially weighted moving average (MA), linearly correlated stellar-activity (SA), moving average with stellar-activity (SAMA), and GP with rotation kernel.}
\end{table*}

\twocolumn
\onecolumn
\section{GJ 699}\label{sec:appendix_gj699}

\begin{table*}[h!]
    \caption{\label{tab:gj699_stats_fwhm}GJ 699 FWHM ESPRESSO model statistics. }
    \centering
    \begin{tabular}{lcccccc}
    \toprule
     Model          &$\ln{P}-$\lnz&\lnz     & \lnzerr   & BIC        & $\chi^2_\nu$ & RMS \\
    \midrule
    \midrule
    WN ($K_1$)      &3.36±0.11   &-354.58±0.11& 0.11&705.65±0.23  &1.07±0.01  &2.23±0.02  \\
    WN ($K_2$)      &10.05±0.13  &-326.64±0.12& 0.12&640.48±1.03  &1.11±0.02  &1.66±0.01  \\
    WN ($K_3$)      &7.13±0.46   &-316.01±0.44& 0.13&627.35±0.34  &1.16±0.01  &1.45±0.01  \\
    \gpgrot ($K_0$) &-10.36±0.62 &-217.46±0.06& 0.09 &445.72±0.31  &0.62±0.01  &0.36±0.01 \\
    \gprot ($K_0$)  &-4.17±0.02  &-215.86±0.01& 0.08 &446.75±0.07  &0.67±0.01  &0.38±0.01 \\
    \gprot ($K_1$)  &-6.61±0.09  &-216.20±0.08& 0.08 &453.24±0.16  &0.76±0.01  &0.41±0.01 \\
    \gprot+Mag(a)   &-11.12±0.99 &-219.33±0.06& 0.10 &449.53±1.98  &0.69±0.04  &0.41±0.01 \\
    \gprot+Mag(b)   &-18.92±0.22 &-219.49±0.08& 0.09 &465.48±0.38  &0.68±0.04  &0.37±0.01 \\
    \bottomrule
    \end{tabular}
    
    \tablefoot{Runs from different models: ($K_n$) denotes the number of Keplerian signals. In descending order: white noise only (WN), Gaussian process with different parameterisations of the rotation kernel \gprot~and \gpgrot, and the inclusion of a magnetic cycle (Mag). This last model was tested with two different boundaries for the magnetic cycle: (a)\Uniform{0.5}{5000}, (b)\Uniform{800}{5000}.}
\end{table*}

\begin{table*}[h!]
    \caption{\label{tab:gj699_stats_fwhm_eh} GJ 699 FWHM ESPRESSO+HARPS model statistics.}
    \centering
    \begin{tabular}{llccccc}
    \toprule
     Model       & $\ln{P}-$\lnz  &\lnz         & \lnzerr  & BIC & $\chi^2_\nu$ & RMS \\
    \midrule
    \midrule
    \gprot ($K_0$)&-3.95±0.11  &-558.34±0.11& 0.11 &1142.38±0.05  &0.84±0.02  &1.27±0.01  \\
    \gprot+Mag(a) &-12.27±0.12 &-559.35±0.07& 0.11 &1158.34±0.37  &0.83±0.02  &1.26±0.01  \\
    \gprot+Mag(b) &-15.51±0.43 &-559.21±0.04& 0.11 &1164.46±0.03  &0.86±0.01  &1.27±0.01  \\
    \bottomrule
    \end{tabular}
    \tablefoot{GP with just a rotation kernel \gprot, and GP with rotation kernel plus a magnetic cycle with different period priors, (a)\Normal{3250}{300}, and (b)\Uniform{800}{5000}.}
\end{table*}

\begin{table*}[h!]
\caption{\label{tab:gj699_stats_rv}GJ 699 RV ESPRESSO model statistics.}
\centering
\begin{tabular}{lcccccc}
\toprule
 Model  & $\ln{P}-$\lnz & \lnz  & \lnzerr & BIC & $\chi^2_\nu$ & RMS \\
\midrule
\midrule
\gpthreesho ($K_0$)  &-28.29±0.01 &-230.28±0.01 &0.11&499.70±0.11 &0.75±0.02 &0.58±0.01 \\
\gpthreesho ($K_1$)  &-13.66±0.30 &-230.84±0.25 &0.26&482.01±0.09 &0.71±0.02 &0.44±0.01 \\
\midrule
\gprotplussho ($K_0$)&-15.70±0.05 &-233.14±0.04 &0.07&495.65±0.22 &0.77±0.01 &0.61±0.01 \\
\gprotplussho ($K_1$)&-2.38±0.54  &-232.56±0.33 &0.09&478.10±0.64 &0.73±0.07 &0.43±0.01 \\
\midrule
\gptworot ($K_0$)    &-19.81±0.05 &-232.25±0.05 &0.07&505.65±0.21 &0.78±0.01 &0.61±0.01 \\
\gptworot ($K_1$)    &-7.54±0.30  &-231.24±0.30& 0.09&489.76±0.12 &0.80±0.03 &0.46±0.01 \\
\midrule
\gprot$+S$ ($K_1$)   &7.55±0.20   &-231.21±0.18& 0.11&470.94±0.25  &0.72±0.03  &0.46±0.01 \\
\gprot+Mag ($K_1$)   &3.67±0.42   &-230.58±0.42& 0.12&480.55±0.62  &0.78±0.01  &0.45±0.01 \\
\midrule

\gpgrot$+S$ ($K_1$)  &4.34±0.11   &-229.86±0.07& 0.10&461.31±0.27  &0.74±0.03  &0.44±0.01  \\
\gpgrot+Mag ($K_1$)  &1.19±0.32   &-230.55±0.28& 0.11&472.50±0.40  &0.70±0.03  &0.44±0.01  \\
\bottomrule
\end{tabular}
\tablefoot{Runs from different models, ($K_n$) denotes the number of Keplerian signals. In descending order: Gaussian process with 3 SHO kernels, rotation plus SHO kernel, 2 rotation kernels, rotation kernel plus a sinusoid, rotation kernel plus magnetic cycle, and the modified rotation kernel.}
\end{table*}

\newpage
\begin{table*}[h!]
\caption{\label{tab:gj699_params_fwhm}GJ 699 FWHM ESPRESSO parameter comparison.}
\centering
\begin{tabular}{llcccccc}
\toprule
Parameter        & Prior            & WN($K_3$)                    & \gpgrot ($K_0$)        & \gprot ($K_0$)              & \gprot ($K_1$)               &\gprot+Mag(a)           &\gprot+Mag(b)              \\
\midrule

\midrule
\multicolumn{8}{c}{Offsets and Jitter} \\
\midrule
$\gamma_{\mathrm{E18}}$&\Uniform{-10.5}{10..5} &$5.25^{+0.96}_{-1.24}$&$1.17^{+0.46}_{-1.29}$&$2.33^{+0.34}_{-1.59}$&$2.75^{+0.72}_{-1.32}$ & $0.14^{+1.74}_{-0.36}$ & $0.62^{+2.39}_{-2.33}$ \\
$\gamma_{\mathrm{E19}}$&\Uniform{-10.5}{10.5} &$0.15^{+0.26}_{-0.13}$&$0.91^{+0.31}_{-0.25}$&$0.89^{+0.38}_{-0.17}$&$0.94^{+0.27}_{-0.13}$ & $1.19^{+0.17}_{-0.19}$ & $0.17^{+1.41}_{-1.72}$ \\
$\sigma_{\mathrm{E18}}$&\Normal{0.5}{1}       &$0.98^{+0.34}_{-0.37}$&$0.54^{+0.15}_{-0.07}$&$0.53^{+0.13}_{-0.08}$&$0.50^{+0.18}_{-0.05}$ & $0.57^{+0.16}_{-0.08}$ & $0.55^{+0.14}_{-0.08}$ \\
$\sigma_{\mathrm{E19}}$&\Normal{0.5}{1}       &$1.43^{+0.10}_{-0.09}$&$0.32^{+0.02}_{-0.03}$&$0.34^{+0.01}_{-0.03}$&$0.36^{+0.02}_{-0.03}$ & $0.38^{+0.01}_{-0.04}$ & $0.37^{+0.04}_{-0.06}$ \\

\midrule

$P_1$            &\Uniform{50}{300} &$165.38^{+0.55}_{-0.11}$& -                         & -                         & $249.73^{+4.40}_{-3.36}$& -                       & -                           \\
$K_1$            &\Uniform{0}{6.60} &  $2.75^{+0.22}_{-0.30}$& -                         & -                         &   $3.57^{+0.52}_{-0.32}$& -                       & -                           \\
$\phi_1$         &\Uniform{0}{2\pi} &  $4.41^{+0.15}_{-0.26}$& -                         & -                         &   $2.54^{+0.63}_{-1.12}$& -                       & -                           \\
$e_1$            &\Normal{0}{0.1}   &  $0.23^{+0.07}_{-0.09}$& -                         & -                         &   $0.02^{+0.03}_{-0.02}$& -                       & -                           \\
$\omega_1$ &\Uniform{0}{2\pi} &  $5.24^{+0.39}_{-0.06}$& -                         & -                         &   $1.28^{+1.29}_{-0.42}$& -                       & -                           \\

$P_2$            &\Uniform{50}{300} &$244.41^{+1.50}_{-1.31}$& -                         & -                         & -                        & -                       & -                           \\
$K_2$            &\Uniform{0}{6.60} &  $4.39^{+0.19}_{-0.01}$& -                         & -                         & -                        & -                       & -                           \\
$\phi_2$         &\Uniform{0}{2\pi} &  $1.02^{+0.40}_{-0.33}$& -                         & -                         & -                        & -                       & -                           \\
$e_2$            &\Normal{0}{0.1}   &  $0.15^{+0.06}_{-0.04}$& -                         & -                         & -                        & -                       & -                           \\
$\omega_2$ &\Uniform{0}{2\pi} &  $2.47^{+0.23}_{-0.29}$& -                         & -                         & -                        & -                       & -                           \\

$P_3$            &\Uniform{1}{5000} & $2384^{+2616}_{-888}$  & -                         & -                         & -                        & -                       & -                           \\
$K_3$            &\Uniform{0}{6.60} &$1.58^{+0.21}_{-0.15}$  & -                         & -                         & -                        & -                       & -                           \\
$\phi_3$         &\Uniform{0}{2\pi} &$5.13^{+1.15}_{-0.15}$  & -                         & -                         & -                        & -                       & -                           \\
$e_3$            &\Normal{0}{0.1}   &$0.05^{+0.06}_{-0.05}$  & -                         & -                         & -                        & -                       & -                           \\
$\omega_3$ &\Uniform{0}{2\pi} &$4.01^{+0.64}_{-1.00}$  & -                         & -                         & -                        & -                       & -                           \\
\midrule

\gpgrot $\rho$   &\Uniform{50}{300} & -                      &$221.98^{+8.55}_{-10.41}$& -                         & -                        & -                       & -                           \\
\gpgrot $\tau$   &\Uniform{50}{600} & -                      &  $28.65^{+1.28}_{-3.11}$& -                         & -                        & -                       & -                           \\
\gpgrot $A_1$    &\Normal{3.3}{3.3} & -                      &   $2.60^{+0.22}_{-0.61}$& -                         & -                        & -                       & -                           \\
\gpgrot $A_2$    &\Normal{3.3}{6.6} & -                      &  $3.70^{+0.21}_{-0.05}$& -                         & -                        & -                       & -                           \\
\midrule
\gprot$\rho$     &\Uniform{50}{300} & -                      & -                         &$170.95^{+22.01}_{-1.98}$ &$151.46^{+10.27}_{-0.81}$ & $155.3^{+2.44}_{-6.04}$ & $156.97^{+100.99}_{-34.51}$ \\
\gprot$\sigma$   &\Normal{3.3}{3.3} & -                      & -                         &   $3.66^{+0.28}_{-0.14}$ &   $2.80^{+0.63}_{-0.52}$ & $2.47^{+0.63}_{-0.51}$  & $3.44^{+0.99}_{-0.59}$      \\
\gprot$Q0$       &\Uniform{0}{100}  & -                      & -                         &   $1.37^{+0.62}_{-0.97}$ &   $4.13^{+2.01}_{-3.60}$ & $2.81^{+2.74}_{-1.72}$  & $1.93^{+0.56}_{-1.56}$      \\
\gprot$\delta Q$ &\Uniform{0}{100}  & -                      & -                         &   $0.20^{+0.13}_{-0.20}$ &   $0.74^{+0.59}_{-0.74}$ & $3.90^{+1.06}_{-0.63}$  & $0.62^{+1.34}_{-0.62}$      \\
\gprot$f$        &\Uniform{0}{1}    & -                      & -                         &   $0.23^{+0.18}_{-0.08}$ &   $0.51^{+0.24}_{-0.28}$ & $0.43^{+0.07}_{-0.18}$  & $0.23^{+0.77}_{-0.23}$      \\
\midrule
$P_{\mathrm{mag}}$    &\Uniform{1}{5000}& -                  & -                         & -                         & -                        & $506.21^{+4.71}_{-8.03}$& $809.26^{+4190.46}_{-9.26}$ \\
$A1_{\mathrm{mag}}$   &\Uniform{0}{6.6} & -                  & -                         & -                         & -                        & $1.07^{+0.53}_{-0.29}$  & $1.47^{+1.36}_{-1.47}$      \\
$A2_{\mathrm{mag}}$   &\Uniform{0}{6.6} & -                  & -                         & -                         & -                        & $0.10^{+0.54}_{-0.10}$  & $1.64^{+1.03}_{-0.57}$      \\
$\phi1_{\mathrm{mag}}$&\Uniform{0}{2\pi}& -                  & -                         & -                         & -                        & $3.75^{+1.04}_{-0.75}$  & $2.80^{+1.27}_{-2.80}$      \\
$\phi2_{\mathrm{mag}}$& \Uniform{0}{2\pi}& -                 & -                         & -                         & -                        & $4.13^{+0.28}_{-0.27}$  & $5.30^{+0.49}_{-1.50}$      \\

\bottomrule
\end{tabular}
\tablefoot{Runs from different models: ($K_n$) denotes the number of Keplerian signals. In descending order: white noise only (WN), Gaussian process with a rotation kernel (\gprot), and GP with a rotation kernel and a magnetic cycle (\gprot+Mag). This last model was tested with two different priors for the magnetic cycle: (a) $\mathcal{U}(0.5, 5000)$, and (b) $\mathcal{U}(800, 5000)$.}
\end{table*}

\begin{table*}
\caption{\label{tab:gj699_params}GJ 699 RV parameter comparison table.}
\centering
\begin{tabular}{lllllll}
\toprule
   Parameter            & Prior         & \gpthreesho        & \gprot S ($K_1$)         & \gpgrot S              & \gptworot              &  \gpgrot+Mag   \\
\midrule
\midrule
\multicolumn{7}{c}{Offsets and Jitter} \\
\midrule
$\gamma_{\mathrm{E18}}$&\Normal{0}{3}   &$-0.03^{+0.32}_{-0.45}$&$-0.39^{+0.28}_{-0.45}$& $-0.52^{+0.72}_{-0.03}$&$-0.20^{+0.55}_{-0.24}$&$-0.40^{+0.61}_{-0.05}$ \\
$\gamma_{\mathrm{E19}}$&\Normal{0}{3}   &$-1.12^{+0.58}_{-0.19}$&$-1.3^{+0.22}_{-0.29}$ & $-1.53^{+0.55}_{-0.59}$& $-0.08^{+0.2}_{-0.40}$&$-1.50^{+0.27}_{-0.20}$ \\
$\sigma_{\mathrm{E18}}$&\LNormal{0.5}{1}& $0.39^{+0.10}_{-0.06}$&$0.55^{+0.2}_{-0.18}$  & $0.45^{+0.05}_{-0.11}$ & $0.45^{+0.10}_{-0.05}$&$0.56^{+0.20}_{-0.20}$  \\
$\sigma_{\mathrm{E19}}$&\LNormal{0.5}{1}& $0.53^{+0.05}_{-0.05}$&$0.55^{+0.04}_{-0.01}$ & $0.54^{+0.03}_{-0.01}$ & $0.55^{+0.04}_{-0.01}$&$0.56^{+0.04}_{-0.01}$  \\

\midrule
\multicolumn{7}{c}{Rotation} \\
\midrule

\gpsho $\rho_1$   & \Uniform{50}{300}   &    $0.37^{+0.02}_{-0.35}$&-                    &-                        &- &-  \\
\gpsho $\sigma_1$ & \LNormal{0.5}{2}    &$145.63^{+29.93}_{-59.63}$&-                    &-                        &- &-  \\
\gpsho $\tau_1$   & \LNormal{3}{2}      &    $19.3^{+4.15}_{-2.46}$&-                    &-                        &- &-  \\
\midrule
\gpsho $\rho_2$   & \Uniform{25}{150}  &    $1.50^{+0.01}_{-0.18}$ &-                    &-                        &- &-  \\
\gpsho $\sigma_2$ & \LNormal{0.5}{2}   &   $64.74^{+5.03}_{-1.43}$ &-                    &-                        &- &-  \\
\gpsho $\tau_2$   & \LNormal{3}{2}     &   $27.93^{+5.33}_{-5.81}$ &-                    &-                        &- &-  \\
\midrule
\gprot$\rho$     &\Uniform{50}{300}    &-                          &$124.68^{+12.14}_{-14.67}$& -                  & $148.27^{+2.69}_{-6.15}$ &-  \\
\gprot$\sigma$   &\Uniform{0}{7.1}     &-                          &$1.51^{+0.24}_{-0.22}$    & -                  &   $2.04^{+0.06}_{-0.21}$ &-  \\
\gprot$Q0$       &\Uniform{0}{50}      &-                          &$1.74^{+1.05}_{-0.89}$    & -                  &   $1.35^{+0.17}_{-0.55}$ &-  \\
\gprot$\delta Q$ &\Uniform{0}{50}      &-                          &$0.28^{+0.05}_{-0.28}$    & -                  & $28.31^{+21.69}_{-14.6}$ &-  \\
\gprot$f$        &\Uniform{0}{1}       &-                          &$0.88^{+0.12}_{-0.01}$    & -                  &   $0.82^{+0.18}_{-0.18}$ &-  \\

\midrule
\gpgrot$\rho$     &\Uniform{50}{300}   &-                           &-                       &$124.7^{+14.04}_{-18.01}$&- &$128.18^{+2.49}_{-4.19}$  \\
\gpgrot$\tau$     &\LNormal{3}{2}      &-                           &-                       &$27.82^{+5.15}_{-5.64}$  &- & $50.62^{+0.11}_{-0.62}$  \\
\gpgrot A1        &\LNormal{0.5}{0.5}  &-                           &-                       &$1.30^{+0.72}_{-0.64}$   &- &   $2.8^{+0.99}_{-0.74}$  \\
\gpgrot A2        &\LNormal{0.5}{0.5}  &-                           &-                       &$3.46^{+1.17}_{-1.24}$   &- &   $6.9^{+0.17}_{-0.41}$  \\

\midrule
\multicolumn{7}{c}{Magnetic Cycle} \\
\midrule

\gpsho $\rho_3$  & \Normal{3250}{300}&    $1.41^{+0.81}_{-0.02}$ &-                          &- &- &-  \\
\gpsho $\sigma_3$& \LNormal{0.5}{0.5}&$3033.9^{+316.9}_{-275.8}$ &-                          &- &- &-  \\
\gpsho $\tau_3$  & \Uniform{0}{3e4}  &$17369^{+12430}_{-8566}$   &-                          &- &- &-  \\
\midrule
$P_m$            & \Normal{3250}{300}&-                          &$2984.0^{+291.1}_{-250.7}$ &$3225.31^{+68.66}_{-176.8}$ &- &$3227.5^{+35.4}_{-196.7}$    \\
$K_m$            & \LNormal{0.5}{2}  &-                          &$2.7^{+0.33}_{-0.29}$      &$3.04^{+0.75}_{-0.83}$      &- &     $2.97^{+0.17}_{-0.38}$  \\
$\phi_m$         & \Uniform{0}{2\pi} &-                          &$4.73^{+0.18}_{-0.18}$     &$4.88^{+0.06}_{-0.11}$      &- &     $4.87^{+0.02}_{-0.11}$  \\
$K_2$            & \LNormal{0.5}{2}  &-                          &-                          &-                         &- &  \\
$\phi_2$         & \Uniform{0}{2\pi} &-                          &-                          &-                         &- &  \\
\midrule
\gprot$\rho$    &\Uniform{800}{5000} &-                          &-                          &-                         &$3186.0^{+159.9}_{-87.4}$&-  \\
\gprot$\sigma$  &\Uniform{0}{7.1}    &-                          &-                          &-                         &$1.28^{+0.34}_{-0.23}$   &-  \\
\gprot$Q0$      &\Uniform{0}{50}     &-                          &-                          &-                         &$28.24^{+19.78}_{-14.27}$&-  \\
\gprot$\delta Q$&\Uniform{0}{50}     &-                          &-                          &-                         &$8.16^{+9.13}_{-8.1}$    &-  \\
\gprot$f$       &\Uniform{0}{1}      &-                          &-                          &-                         &$0.65^{+0.33}_{-0.28}$   &-  \\

\midrule
\multicolumn{7}{c}{Keplerian Orbit} \\
\midrule

$P_1$            & \Uniform{0.5}{50} &$3.1536^{+5e-5}_{-5e-4}$&$3.1529^{+7e-4}_{-5e-4}$&$3.15^{+6e-5}_{-5e-4}$ &$3.1534^{+2e-4}_{-2e-4}$&$3.1535^{+3e-4}_{-4e-4}$\\
$K_1$            & \Uniform{0}{6.5}  &$0.59^{+0.08}_{-0.09}$  &$0.56^{+0.01}_{-0.08}$  &$0.56^{+0.03}_{-0.04}$ &$0.55^{+0.04}_{-0.03}$  &$0.57^{+0.08}_{-0.08}$  \\
$\phi_1$         & \Uniform{0}{2\pi} & $3.95^{+1.68}_{-2.3}$  &$4.21^{+1.68}_{-2.56}$  &$1.53^{+1.54}_{-0.15}$ &$2.44^{+1.07}_{-0.63}$  &$1.27^{+1.68}_{-0.05}$  \\
$e_1$            & \Normal{0}{0.1}   &$0.03^{+0.01}_{-0.03}$  &$0.04^{+0.05}_{-0.04}$  &$0.03^{+0.01}_{-0.03}$ &$0.01^{+0.02}_{-0.01}$  &$0.06^{+0.05}_{-0.06}$  \\
$\omega_1$ & \Uniform{0}{2\pi} &$4.22^{+2.06}_{-0.03}$  & $3.7^{+1.76}_{-2.26}$  &$0.48^{+2.63}_{-0.48}$ &$5.58^{+0.7}_{-0.94}$   &$0.67^{+1.82}_{-0.67}$  \\

\bottomrule
\end{tabular}
\end{table*}

\twocolumn

\end{appendix}

\end{document}